\documentclass[prd,aps,showpacs,
nofootinbib,%
notitlepage,
eqsecnum,%
twocolumn
]{revtex4-2}

\usepackage{graphicx}
\usepackage{amsmath}
\usepackage{amssymb}
\usepackage{hyperref} 
\usepackage{upgreek}
\usepackage{comment}

\usepackage{color}

\newcommand{\new}[1]{{\color{blue} #1}} 

\renewcommand{\new}[1]{#1} 

\newcommand{\be}{\begin{equation}}
\newcommand{\ee}{\end{equation}}
\newcommand{\bea}{\begin{eqnarray}}
\newcommand{\eea}{\end{eqnarray}}

\newcommand{\newchange}[1]{{\color{black} #1}}

\newcommand{\J}{\mathcal{J}}

\newcommand{\Q}{\mathcal{Q}}
\newcommand{\mV}{M} 
\newcommand{\mA}{M^A}
\newcommand{\Ai}{\operatorname{Ai}}
\newcommand\Jpq{\widetilde{\mathcal{J}}}

\DeclareMathOperator{\asinh}{arsinh}

\begin{document}

\title{Divergences 
in the hadronic light-by-light amplitude of\\
the holographic soft-wall model}

\author{Josef Leutgeb}
\author{Jonas Mager}
\author{Anton Rebhan}
\affiliation{Institut f\"ur Theoretische Physik, Technische Universit\"at Wien,
        Wiedner Hauptstrasse 8-10, A-1040 Vienna, Austria}

\date{\today}

\begin{abstract}
We use the {Wentzel-Kramers-Brillouin (WKB)} approximation to uncover divergences and instances of non-commuting limits in a large class of holographic soft-wall models. We show that the infinite sum over single resonance contributions for a variety of observables involving the Chern-Simons term, such as the {vector-vector-axial-vector (VVA)} correlator or the hadronic light-by-light tensor, does not converge. These divergences can in some cases (such as the VVA correlator) be traced back to non-commuting limits and avoided by working directly in the 5-dimensional setup with bulk-to-boundary propagators. However, the hadronic light-by-light scattering tensor also diverges in the 5-dimensional formulation with corresponding Green functions, 
preventing a correct implementation
of the Melnikov-Vainshtein short-distance constraint and even leading to an infinite contribution to the muon $g-2$. 
We also discuss modifications of the standard soft-wall model that are able to resolve this issue.
\end{abstract}

\maketitle

\section{Introduction}\label{sec:intro}

The holographic principle \cite{tHooft:1993dmi,Susskind:1994vu} in its modern form relates strongly coupled gauge theories in $d$ spacetime dimensions to weakly coupled theories of gravity in higher dimensions. 
Shortly after the first discoveries of explicit examples \cite{Maldacena:1997re} of such dualities, there were attempts at constructing a corresponding gravity dual to the theory of the strong interactions \cite{Witten:1998zw,Polchinski:2000uf,Aharony:2002up,Karch:2002sh}.
Although to this date there is no rigorously known holographic dual to QCD, many models have been constructed that capture the essential features of QCD rather well.

\new{Top-down} constructions such as the Witten-Sakai-Sugimoto \cite{Sakai:2004cn,Sakai:2005yt} model
\new{for the low-energy limit of large-$N_c$ QCD}
follow from certain brane configurations in string theory. The general lessons learned in these more rigorous approaches were used to construct bottom-up dual models by hand \cite{Erlich:2005qh,DaRold:2005mxj,Hirn:2005nr}. 
For phenomenological applications these bottom-up models often prove to be more suitable as they
offer the possibility to match to some extent to the 
\new{leading-order}
perturbative limit of QCD at high energies.

A particularly interesting class of models of holographic QCD (hQCD) are so-called soft-wall (SW) models \cite{Karch:2006pv}, which have a particular type of dilaton profile $e^{- \phi}=e^{-c^2z^2}$ \new{on top of an anti-de Sitter (AdS) metric with holographic coordinate $z$.} This breaks conformal invariance, implements confinement, and gives rise to a Regge-like mass spectrum 
\new{for vector mesons with} $\mV_n^2 \sim n$. 
The latter feature is missing in the simple hard-wall (HW) AdS/QCD models, which achieve confinement via a hard cut-off at Poincaré coordinate $z=z_0$ where boundary conditions have to be specified.
The latter have in recent years been very successfully applied \cite{Cappiello:2010uy,Leutgeb:2019zpq,Cappiello:2019hwh,Leutgeb:2019gbz,Leutgeb:2022lqw,Cappiello:2021vzi,Leutgeb:2024rfs,Cappiello:2025fyf,Mager:2025pvz}
to hadronic contributions to the muon $g-2$ \cite{Aliberti:2025beg}, namely the hadronic light-by-light (HLbL) contribution
for which it provided the first hadronic model that implemented the
\new{Melnikov-Vainhstein} short-distance constraint (MV-SDC) \cite{Melnikov:2003xd,Colangelo:2019lpu,Bijnens:2020xnl}
consistently in the chiral limit. Moreover, the quantitative prediction for the contribution
of axial-vector mesons that are crucial for this feature has recently been confirmed remarkably well
by a data-driven dispersive approach \cite{Hoferichter:2024bae,Hoferichter:2024vbu}.

In the present paper we show that the soft-wall model,
which has also been used to evaluate HLbL contributions to the muon $g-2$
\cite{Colangelo:2023een,Colangelo:2024xfh},
is inadequate in this regard. 
In the original soft-wall model, as well as in a large class of deformations of it,
we find
divergences in a variety of observables that in some way or another involve the Chern-Simons term.  
Representing quantities such as the vector-vector-axial (VVA) correlator 
and the HLbL tensor as an infinite sum over modes, we show that the sum cannot converge. 
A more natural way of calculating such observables is by using the 5-dimensional formulation using bulk-to-boundary and bulk-to-bulk propagators. In the case of the VVA correlator, this avoids the divergence, while in the case of the HLbL tensor it does not.
This implies that a large class of soft-wall models cannot be used to provide input to the Standard Model calculation of the muon $g-2$. Contributions from \emph{individual} intermediate states as calculated recently in \cite{Colangelo:2023een,Colangelo:2024xfh}
are well defined; however, the summation over all such states, \new{which is
necessary to satisfy short-distance constraints of the HLbL amplitude,}
diverges.

\newchange{A crucial result is the derivation of a lower bound on the magnitude of the transition form factor (TFF) of highly excited pseudoscalars. With mass $m_n\to\infty$, we obtain the exponential behavior
\begin{align}
\label{eq:expscalsimple}
      |F_n(Q_1^2,Q_2^2)| \gtrsim   \; \tilde{C} \;m_n^{\lambda} e^{\frac{m_\rho^2}{16\pi  \sigma} m_n}
\end{align}}
for fixed $Q_i$, where 
$\tilde{C}$ and
$\lambda=-\frac{Q_1^2+Q_2^2}{\new{m_\rho^2}}-\frac{1}{4}$ depend on $Q_i$, but are independent of $m_n\sim n^{2/3}$. \new{($\sigma$ is the chiral condensate.)
A similar result is obtained for the TFF of excited
axial vector mesons.}
This exponential scaling of the TFFs is the reason for \new{various} convergence problems; it is derived in detail in section \ref{sec:mainexpscal} using Wentzel–Kramers–Brillouin (WKB) techniques
after a brief review of the soft-wall models in section \ref{sec:setup}.

In section \ref{sec:noncomm}, we investigate the VVA correlator and show that when computed via summation over single particle poles, a divergent expression results. We then show that this divergence may be circumvented by working directly in the 5-dimensional setup. It will become clear that the divergence is a result of illegally exchanging an infinite sum with an integral over an infinite region. A similar situation occurs for various pseudoscalar sum rules.

A similarly non-convergent sum is observed for the HLbL tensor in section \ref{sec:Lightbylighttensor}. However, in this case a divergence is unavoidable and it appears in the 5-dimensional formulation as well. 

Finally, in section \ref{sec:varsandkiritsis}, we extend our analysis to a much larger set of bulk theories by considering general warp factors, dilaton profiles, and bi-fundamental background solutions $X_0$, only assuming a Regge spectrum of \new{vector meson} masses. We then comment on modifications of the model which avoid the pathological behavior \eqref{eq:expscalsimple} of the transition form factor. The simplest of such modifications seems to be either a hard-wall cutoff or a scalar extension of the Chern-Simons term, which is obtained when combining the bifundamental scalar $X$ and the gauge fields into a superconnection \new{as done
in the V-QCD model of Refs.\
\cite{Casero:2007ae,Gursoy:2007cb,Gursoy:2007er,Iatrakis:2010zf,Iatrakis:2010jb,Jarvinen:2011qe}.
Without such an extension of the Chern-Simons term, the alternative soft-wall models considered in \cite{Batell:2008zm,Gherghetta:2009ac,Kapusta:2010mf} do not even
have finite transition form factors at all. They are rescued by having
an unbounded field $X_0$ in accordance with Ref.~\cite{Shifman:2007xn}, 
in contrast to the dynamical soft-wall model in
\cite{Li:2012ay,Chen:2022goa}, where chiral symmetry is completely restored
for highly excited mesons.
The problem revealed by our analysis thus leads to strong constraints on
holographic QCD model building.
}

\section{Pseudoscalar and axial sector in the soft-wall model}
\label{sec:setup}

In this section, we first present the general setup for soft-wall models. When applying the WKB approximation in the pseudoscalar and axial vector sector, we first specialize to the specific version of the model used in \cite{Colangelo:2023een,Colangelo:2024xfh}.
The holographic soft-wall model contains\footnote{The group of 0-form flavor symmetries in QCD is more accurately given by $\frac{U(N_f)_L \times U(N_f)_R}{\mathbb{Z}_{N_c}}$ (not accounting for the $U(1)_A$ anomaly). The division by $\mathbb{Z}_{N_c}$ happens due to gauge transformations. The global part of the symmetry group plays no role in the rest of this paper.} $U(N_f)_L \times U(N_f)_R$ gauge fields $(A_L,A_R)$ and a complex bifundamental scalar $X$, which are dual to the chiral flavor currents $(J_L,J_R)$ and chiral condensate $\psi_L^{\dagger} \psi_R$, respectively. Additionally, there is the 5-dimensional metric $g$ and a dilaton $e^{-\phi}$ \new{corresponding to} the 4-dimensional energy-momentum tensor and the Lagrangian. The action of the soft-wall model reads
\begin{align}
\label{eq:SWaction}
    S=&  -\frac{1}{4g_5^2}\!\!\int\limits_{AdS_5} \!\!\!\sqrt{ g } e^{-\phi} \,\text{tr} \Bigl[(F_L)_{MN}(F_L)^{MN} \nonumber\\
    & \qquad\qquad\qquad\qquad + (F_R)_{MN}(F_R)^{MN}\Bigr] \nonumber \\&+\int\limits_{AdS_5} \!\!\! \sqrt{g } e^{-\phi} \,\text{tr} \left[D_M X D_N X^{\dagger}g^{M N}+ 3X X^{\dagger}\right] \nonumber\\
    & +S_{\text{CS}}[A_L,A_R] \nonumber \\
    &-2 k_T \int \sqrt{g} e^{-2 \phi}(R-2 \Lambda)
\end{align}
with a Chern-Simons term $S_\mathrm{CS}[A_L,A_R]=S_\mathrm{CS}^\mathrm{L}-S_\mathrm{CS}^\mathrm{R}$, which is independent of the dilaton and which accounts for the flavor anomalies of QCD
\be\label{SCS}
S_{\rm CS}^\mathrm{L,R}=\frac{N_c}{24\pi^2}\int\text{tr}\left(\mathcal{B}\mathcal{F}^2-\frac{i}2 \mathcal{B}^3\mathcal{F}
-\frac1{10}\mathcal{B}^5\right)^\mathrm{L,R}.
\ee
\new{The Chern-Simons term plays a key role in the determination of TFFs for pseudoscalars and axial vectors.}
Within the realm of effective field theory, given the field content and the chiral anomaly of QCD, this action can be \new{motivated as containing} the most relevant terms in a derivative expansion. The mass of the scalar $m_X^2=-3$ models the UV scaling dimension of the quark condensate operator in QCD.

In this section we take the gravitational background to be $\mathrm{AdS}_5$ whose metric in Poincar\'e coordinates is 
\be
ds^2= w^2(z)(\eta_{\mu \nu}dx^\mu dx^{\nu}-dz^2),
\quad w(z)=\frac1z. 
\ee
Assuming the quadratic dilation profile $\phi=c^2 z^2$ \new{without modifying the background otherwise, one obtains a}
Regge behavior $\mV_n^2\sim n$ for the masses of scalar and vector mesons.

The bi-fundamental scalar field $X$ obtains a vacuum expectation value $X_0$. This solution goes like $\frac{1}{2}(m_q z+\sigma z^3)$ close to $z=0$ and is the sum of a constant and $e^{z^2}$, an exponentially growing function, at $z= \infty$. In order to have a regular solution everywhere (not just on the Poincaré patch), we need $X_0$ to go to a constant, which would imply a linear relation between $m_q$ and $\sigma$. This is, of course, not realized in QCD, which spontaneously breaks the chiral symmetry at $m_q=0$. As described in \cite{Karch:2006pv,Kwee:2007nq} one can add a general potential $U(X X^{\dagger})$ to the action to engineer a physically more sensible regular background $X_0$.

It is also possible to consider singular solutions which go to infinity as $z\rightarrow \infty$ (see for example \cite{Gubser:2000nd} for singularities in AdS/CFT). In principle then one really has to calculate backreactions of the scalar vacuum expectation value to the metric (which could be neglected for regular backgrounds for $N_c \gg N_f$), and we will comment on models where this happens such as V-QCD in later sections. For the first part of the paper we will however simply follow \cite{Colangelo:2023een,Colangelo:2024xfh} and \new{set} 
\be
X_0=\frac{1}{2}v(z)=\frac{1}{2}(m_q z+\sigma z^3)
\ee
for all $z$ without backreactions to the metric, as this is the most commonly used form in the literature. After understanding the calculations with this background, the analysis can be readily generalized to more general $X_0$, warp factor $w(z)$, and dilaton profile, which will be analyzed in section \ref{sec:varsandkiritsis}. As will be seen there, when $X_0 \rightarrow \text{const.}$ at infinity, there will be even problems with the definition of individual transition form factors. \new{Restricting ourselves to the $N_f=2$,} the parameters $c$, $\sigma$, and $m_q=m_{u,d}$  are fitted to the $\rho$-meson mass ($m_\rho=2c$), the pion decay constant and mass, respectively, while $g_5=2 \pi$ follows from a fit of the vector-current correlator to the OPE of QCD.

We parametrize pseudoscalar and axial vector fluctuations of the dynamical fields by $X=e^{i\eta}X_0 e^{i \eta}, \; A_{\mu}=[(A_L)_{\mu}-(A_R)_{\mu}]/2=\partial_{\mu} \varphi+ A^{\perp}_{\mu}$ with $\partial^{\mu}A_{\mu}^\perp=0$. For the purposes of this paper, it is sufficient to consider a uniform $X_0 \sim \mathbb{I}$. For the most part we will even work in the chiral limit; a nonzero quark mass is only needed for the discussion of the pseudoscalar decay constants.
In the $(A_L)_z=(A_R)_z=0$ gauge, the \new{holographic} equations of motion for pseudoscalar fluctuations \new{with mass $m_n$} following from the action \eqref{eq:SWaction} read
\begin{align}
\label{eq:pseq}
     \partial_z(\frac{e^{-\phi}}{z}\partial_z\varphi)+g_5^2 \frac{e^{-\phi}}{z^3}v^2(\eta-\varphi)=0\\
    \frac{m_n^2}{g_5^2} \frac{\partial_z \varphi}{z}= \frac{v^2}{z^3}\partial_z \eta,
\end{align}
while for the axial sector they read
\begin{align}
\label{eq:axeq}
    \partial_z(\frac{e^{-\phi}}{z}\partial_z A^{\perp}_{\mu}) + \frac{(M_n^A)^2}{z} e^{-\phi}A^{\perp}_{\mu}-g_5^2 \frac{v^2}{z^3}e^{-\phi} A^{\perp}_{\mu}=0.
\end{align}

The normalizable modes obey $\varphi(\infty)=\eta(\infty)$ and $\partial_z\varphi(\infty)=\partial_z\eta(\infty)=0$ for the pseudoscalars and $A^{\perp}(\infty)=\partial_zA^{\perp}(\infty)=0$ for the axial vectors'
\new{holographic wave function in $A_\mu^\perp(x,z)=A_\mu(x)A^\perp(z)$.}

Both the pseudoscalar and the axial-vector equations of motion may be converted into a Schrödinger equation.

In the pseudoscalar sector, we define 
\be
y=\sqrt{\omega}\frac{e^{-\phi}}{z}\partial_z \varphi, 
\qquad \omega={z^3}/[v^2g_5^2 e^{- \phi}], 
\ee
which \new{in the chiral limit} obeys
\begin{align}
     -y'' +V_P(z) y=m_n^2y\equiv E_n y, \\
     V_P(z)= -2c^2+\frac{15}{4 z^2}+c^4 z^2+g_5^2 \sigma^2 z^4,
\end{align}
where $E_n=m_n^2$ has dimension of energy squared. 
The asymptotic form of the potential is governed by the background of the scalar $X_0$.
The normalization condition for the function $y$ is given by
\begin{align}
\label{eq:normconps}
    \frac{m_n^2}{g_5^2}\int_0^\infty y^2 \, dz = 1.
\end{align}
In the axial sector, we define 
\be
\xi = \sqrt{\rho}A^{\perp},
\qquad
\rho =\frac{e^{-\phi}}{z},
\ee
which also leads to a Schrödinger equation, now with a potential (in the chiral limit)
\begin{align}
    V_A(z)= \frac{3}{4 z^2}+c^4z^2+g_5^2 \sigma^2 z^4.
\end{align}
\new{The normalization condition} for the axial-vector \new{mode functions} is 
\begin{align}
\label{eq:axnormcon}
      \frac{1}{g_5^2}\int_0^\infty \xi^2 \, dz =1.
\end{align}

 Since the positive sign of the \new{chosen dilaton profile $\phi=c^2 z^2$} will play an essential role in our derivations, we briefly recall the argument \cite{Karch:2010eg} against a \new{reversed-sign} dilaton \cite{Brodsky:2006uqa,deTeramond:2009xk,Zuo:2009dz}.

Let us calculate the vector current two-point function $\Pi^{\mu \nu}(q)$ for a dilaton profile which at very large $z$ is given by $e^{-\phi} =e^{-a z^2}$, but where the sign of $a$ is not fixed yet. It will be seen that if the sign of $a$ is negative, a massless scalar multiplet appears, signaling the breakdown of the vector flavor symmetry.
In the $V_z=0$ gauge, one may decompose the Fourier transformed vector gauge field $\tilde{V}_{\mu}(z,q)$ as
\begin{align}
    \tilde{V}_{\mu}(z,q)=\J(z,Q) P_{\mu}^{\;\nu}(q)V^{(0)}_{\nu}(q)+q_{\mu}\frac{q^{\nu}V^{(0)}_{\nu}(q)}{q^2},
\end{align}
with $V^{(0)}$ being the boundary value
and $P^{\mu \nu}(q)=\eta^{\mu \nu} -\frac{q^{\mu} q^\nu}{q^2}$. 
The bulk-to-boundary propagator $\J(z,Q)$ 
with virtuality $Q\equiv\sqrt{-q^2}$
obeys the equation
\begin{align}
    \partial_z \left( \frac{e^{-\phi}}{z}\partial_z\J(z,Q)\right)+\frac{q^2 e^{-\phi}}{z}\J(z,Q)=0,
\end{align}
and $\J(0,Q)=1$.
The two-point function \new{is} given by
\begin{align}
    \Pi_{ab}^{\mu \nu}(q)=\frac{1}{g_5^2} \delta_{ab} P^{\mu \nu}(q) \frac{\partial_z \J(z,Q)}{z}\bigg|_{z\to0}.
\end{align}
In order to avoid a pole at $q^2=0$ \new{of the form $\Pi^{\mu\nu}_{ab}\sim \delta_{ab}q^\mu q^\nu/q^2$ from massless scalars in $V_\mu\propto \partial_\mu S$, one needs}
$z^{-1}{\partial_z \J(z,0)}|_{z\to0}$ to vanish\footnote{The familiar divergence ${\partial_z \J(\varepsilon,Q)}/{\varepsilon} \sim q^2 \log (\varepsilon Q)$, which is canceled by holographic renormalization, is not relevant in this discussion as it only appears at nonzero $q^2$. }.
For $a<0$, the most general regular solution of the EOM obeying the UV boundary conditions with $q^2=0$ is
\begin{align}
    \J(z,0)= 1+c_2 \int_0^z dz' \; z' e^{\phi(z')}.
\end{align}
For $a>0$, the second term needs to vanish since it would grow at large $z$, but for $a<0$, any $c_2$ seems to be allowed.
In order to determine $c_2$, we look at small but nonzero values of $q^2$ and demand that $\J(z,Q)$ is regular as $q^2 \rightarrow 0$.
It is then immediately seen that when expanding $\J(z,Q)= \J(z,0)+q^2 \J^{(1)}(z,Q)+...$ the perturbation $\J^{(1)}$ grows logarithmically as $z\rightarrow 0$ if $1+c_2 \int_0^\infty dz\,z\,e^\phi \neq 0$.
Demanding regularity means that
\begin{align}
  \bigg(1+c_2 \int_0^\infty dz\,z\,e^\phi \bigg) = 0 ,
\end{align}
which defines a nonzero value for $c_2$ in the case $a<0$.
Using then that ${e^{-\phi}\partial_z \J(z,0)}/{z}$ is constant in $z$, we may evaluate it at large $z$, where we know the form of the dilaton, which gives
\begin{align}
    \frac{\partial_z \J(z,0)}{z}\bigg|_{z=0}=c_2 \neq 0.
\end{align}
A multiplet of massless scalars is therefore unavoidable with a \new{reversed-sign  ($a<0$)} dilaton.
This line of argumentation may even be generalized further to warp factors different from $1/z$ and non-exponential dilatons $e^{-\phi}$. The authors of \cite{Gutsche:2011vb} have also shown that the pion electromagnetic form factor diverges for the \new{reversed-sign} dilaton.
We will henceforth restrict to $e^{-\phi}=e^{-c^2 z^2}$ for the bulk of the paper. In section \ref{sec:varsandkiritsis} we consider rather general dilation and warp factors with the only constraints coming from demanding 
Regge behavior $m_n^2 \sim n$ \new{for large $n$} and no vector-symmetry breaking.

\section{WKB, transition form factors and hadronic light-by-light scattering}
\label{sec:mainexpscal}

\subsection{WKB and definition of four regions}
The WKB approximation is a powerful method for approximating solutions to differential equations of the form
\begin{align}
     y''+k^2(z)y=0
\end{align}
when $\frac{k'}{k^2}\ll1$\newchange{, where $k' \equiv \partial_z k$}. As shown in the previous section, the mode equations for the pseudoscalar and axial vector sector can be brought into this form with 
\be
k^2(z)=E_n-V_{P,A}(z),\quad E_n=m_n^2
\;\;\text{or}\;\; (M^A_n)^2.
\ee
For generic $z$, $\frac{k'}{k^2}\rightarrow 0$ as $E_n \rightarrow\infty$, so the WKB approximation becomes more and more accurate, the larger $E_n$. The explicit analytic form of the WKB approximation will ultimately allow us to make rigorous statements at large $E_n
$ and prove the assertions stated in the introduction.
When $k^2>0$, the solution within the WKB approximation is oscillating and reads
\begin{align}
    y=\frac{1}{\sqrt{k(z)}}(A\, e^{i\int k}+B\, e^{-i\int k}).
\end{align}
If $k^2=- \kappa^2$ is negative, the solution \new{has exponential behavior}
\begin{align}
     y=\frac{1}{\sqrt{\kappa(z)}}(A\, e^{\int \kappa}+B\, e^{-\int \kappa}).
\end{align}
However, even at very large $E_n$, there will always be regions where $\frac{k'}{k^2}\ll1$ is not obeyed, and we need to find different ways to approximate the solution. These regions are the small-$z$ region, where $V_{P,A}(z)\gg E_n$ and therefore $\frac{k'}{k^2}\sim \frac{1}{z^4}$ (the size of this region shrinks, as $E_n$ grows), and the regions around the points $z_l,z_r$ where $k^2(z)=0$.
For $V_P$, the points $z_l,z_r$ are approximately given by
\begin{align}
    z_l&=\sqrt{\frac{15}{4 E_n}},\\
    z_r&=\bigg(\frac{E_n}{g_5^2 \sigma^2}\bigg)^{\frac{1}{4}},
    \label{eq:zr}
\end{align}
at large $E_n$.

\begin{figure}[t]
    \centering
    \includegraphics[width=1\linewidth]{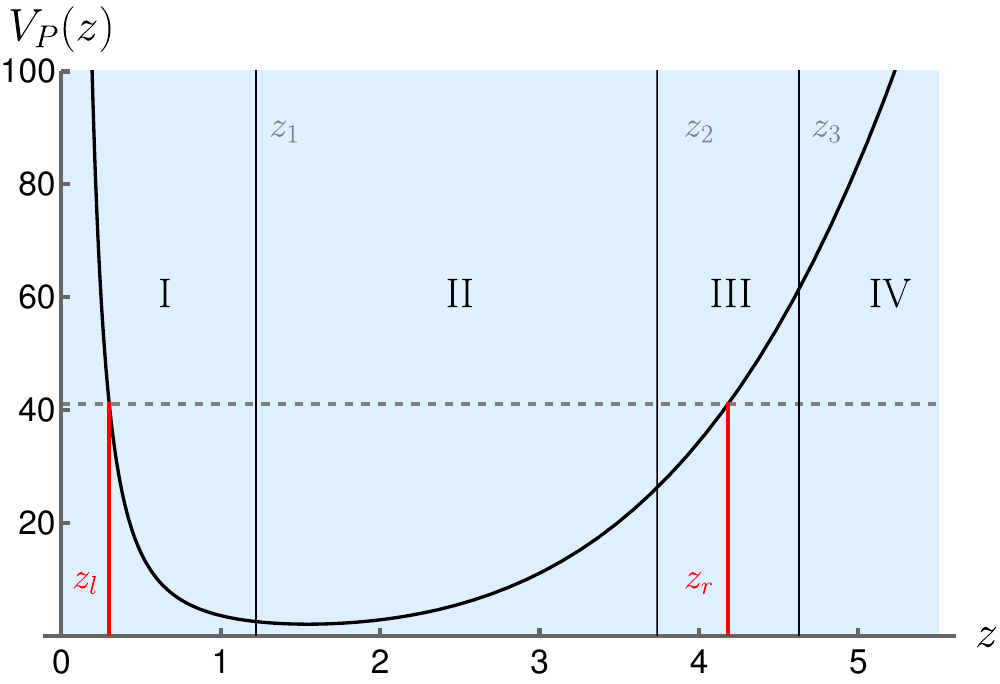}
    \caption{The pseudoscalar potential in units of GeV$^2$ and the four regions into which the whole range of the $z$ coordinate (in units of GeV$^{-1}$) is divided for $E_n\sim 40 $ GeV$^2$, which is indicated by the dashed line. The two locations where the dashed line intersects $V_P(z)$ are the points $z_l,z_r$ where $k^2=0$. The three lines separating the regions are located at $z_1,z_2,z_3$. In regions II and IV, the oscillating and exponentially decaying WKB solutions, respectively, are appropriate. In region I we use a Bessel function, while in region III, an Airy function is used.}
    \label{fig:regions}
\end{figure}

This divides the whole range of the holographic coordinate into four regions\footnote{The point $z_l$ always lies in region I, where we will not use the WKB approximation anyway.} separated by $z_1,z_2,z_3$ as shown in fig. \ref{fig:regions}.
The points $z_i$ separating the regions are chosen in such a way that the solutions on both sides are valid and become exact as $E_n \rightarrow \infty$. We will state their precise definition later and elaborate on the matching in appendix \ref{app:WKBDetails} .

In region I, $z \in [0,z_1]$ the potential is $V_P \sim \frac{15}{4z^2}$ and the equations of motion (EOM) can be solved analytically by
\begin{align}
\label{eq:BesselJtext}
  y(z)=C_1 \sqrt{z}J_2(\sqrt{E_n}z),
\end{align}
where $J_2$ is a Bessel function of the first kind.

In region II, $z \in  [z_1,z_2]$, $k^2>0$ and we can use the oscillatory WKB solution
\begin{align}
\label{eq:WKBeqtext}
      y(z)&=\frac{1}{\sqrt{k(z)}} C_2 \exp \left(i \int_{z_l}^z k(z') dz'\right)\nonumber \\&+ \frac{1}{\sqrt{k(z)}}D_2 \exp \left(-i \int_{z_l}^z k(z') dz'\right) .
\end{align}

In region III, $z \in[z_2,z_3]$, a solution in terms of the Airy function is appropriate
\begin{align}
\label{eq:Airy III text}
      y(z)=& C_3 \Ai\left(L^{\frac{1}{3}}(z-z_r)\right),
\end{align}
where $L=V'|_{z=z_r}$

Finally in region IV, $z \in [z_3,\infty]$, $k^2=-\kappa^2$, we may use the exponentially decaying WKB solution
\begin{align}
        y(z)=C_4  \frac{1}{\sqrt{\kappa}}\exp \left[ - \int_{z_r}^z \kappa(z') dz' \right]
\end{align}
We find the matching conditions \eqref{eq:coefstart}-\eqref{eq:coefend} by demanding that the solutions can be glued together smoothly at $z_i$ as $E_n\rightarrow \infty$. Additionally, one finds a constraint on the allowed energies $E_n$
\begin{align}
\label{eq:energypscond}
     \int_{z_l}^{z_r}k(z)\,dz- \frac{\pi}{2}+ \frac{\sqrt{15}}{4}\pi=n \pi.
\end{align}
The matching conditions fix all but an overall constant which we take to be $C_1$. It will be determined by imposing the \new{normalization} condition \eqref{eq:normconps}.
At large $E_n$, the energies allowed by \eqref{eq:energypscond} obey
\begin{align}
    \frac{2}{3}\frac{E_n^{3/4}}{\sqrt{g_5 \sigma}}K(-1)- \frac{\pi}{2}+ \frac{\sqrt{15}}{4}\pi=n \pi,
\end{align}
where $K(x)$ is the complete elliptic integral of the first kind, implying a non-Regge spectrum $m_n =\sqrt{E_n} \sim n^{2/3}$ for excited pseudoscalars.

The points $z_i$ are chosen to scale with $E_n$ in the following way in order to ensure the validity of the solutions on both sides at $z_i$:
\begin{align}
    z_1&= z_l \left( \frac{E_n}{c^2} \right)^{a_1}, \\
    z_2 &= z_r\left(1-\left( \frac{E_n}{c^2} \right)^{-a_2} \right),\\
    z_3&=  z_r\left(1+\left( \frac{E_n}{c^2} \right)^{-a_2} \right),
\end{align}
with
\begin{align}
    0<a_i <\frac{1}{2}, \quad i=1,2.
\end{align}
It is shown in appendix \ref{app:WKBDetails} that with $z_i$ of this form we obey the criterion of validity for both solutions at $z_i$ at large $E_n$.
There is a whole range of allowed exponents $a_i$, and each choice results in a different $z_i$. Which choice\footnote{One could in principle also introduce numerical constants in front of $\frac{E_n}{c^2}$ and gain even more freedom in the $z_i$. Also this choice will not matter \new{for the following}.} one ultimately makes is irrelevant for all the following results of this paper, but numerically, some choices might be more advantageous, for example, to obtain better agreement at lower $E_n$. One may, in principle, even make the exponents $a_i$ depend on $E_n$ as long as they remain inside the allowed interval. Provided that $\lim_{E_n\rightarrow \infty} a_i(E_n)$ exists and is not equal to $0$ or $\frac{1}{2}$, the gluing procedure will work.

\begin{figure}[t]
    \centering
    \includegraphics[width=1\linewidth]{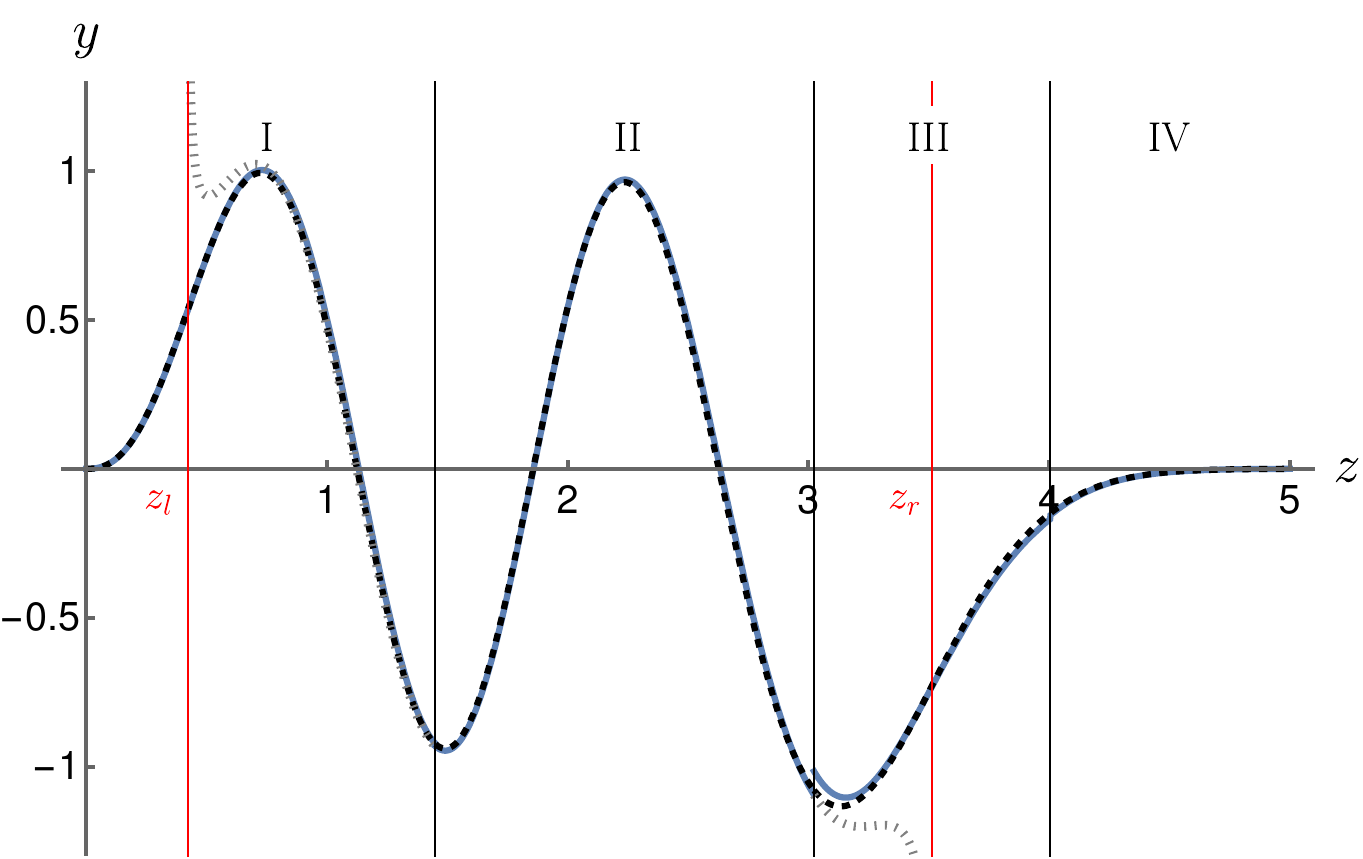}
    \caption{A comparison of the WKB mode function (blue) and the full numerical solution  (dashed black) as a function of $z$ {(in units of $\rm GeV^{-1}$)}. The vertical black lines indicate the boundaries $z_i$ of the four regions, and the dotted gray curve indicates the WKB solution \eqref{eq:WKBeqtext}. Clearly the points $z_l,z_r $ marked by red lines where $\frac{1}{\sqrt{k}}$ diverges are outside region \textrm{II}. }
    \label{fig:comparison}
\end{figure}

As shown in appendix \ref{app:WKBDetails}, the normalization condition fixes $C_1$ which asymptotically obeys
\begin{align}
\label{eq:normconpstext}
    \frac{K(-1)}{\pi g_5^2} \frac{1}{\sqrt{g_5 \sigma}}  C_1^2 E_n^{3/4} =1.
\end{align}
We have therefore fully determined the solution for each of the four regions.
In figure \ref{fig:comparison} we compare the full numerical solution of 
\new{the fourth excited mode with mass about 4.5 GeV} 
to the WKB result. The remarkable agreement already at such \new{low mode number} only improves when going to higher $n$.

In appendix \ref{app:WKBDetails} we have spelled out explicitly also the analysis for the axial vector mesons, which differs from the pseudoscalar case in a rather minimal way. The potential $V_A$ is slightly different at small $z$, \new{leading also to
a mass spectrum $\mA_n\sim n^{2/3}$}, whereas the normalization condition \eqref{eq:axnormcon} does not involve $E_n=(\mA_n)^2$.

Departing from the chiral limit, the pseudoscalar potential changes rather drastically at small $z$, prompting a corresponding change in the solution in region I, which is shown in the appendix. To leading order in $E_n$ the solutions in the other regions stay the same, but the change in region I, now results in a nonzero decay constant for $n\neq0$
\begin{align}
    f_{n}= -\frac{m_q}{g_5}\frac{1}{\sqrt{z}}y\big|_{z=0}= -\frac{m_q }{g_5}C_1,
\end{align}
where \new{for large $n$}
$C_1$ is still given by \eqref{eq:normconpstext}.

\subsection{Transition form factors, sum rules and hadronic light-by-light tensor}
\label{sec:TFFexpscalsubsection}

In this subsection, we calculate the transition form factors of charge-neutral mesons following from the Chern-Simons term. 
In holographic QCD models with asymptotic AdS geometry,
they turn out to reproduce the perturbative QCD results from
light-cone expansions \cite{Lepage:1979zb,*Lepage:1980fj,*Brodsky:1981rp,Efremov:1979qk,Hoferichter:2020lap} for pseudoscalars \cite{Grigoryan:2008up,Brodsky:2011xx,Leutgeb:2021mpu}
as well as axial vector mesons \cite{Leutgeb:2019gbz}.
However, in soft-wall models we find an exponential
scaling in $E_n$ leading to \eqref{eq:expscalsimple}.
This then
implies convergence problems for a variety of infinite sums. Among these quantities are sum rules involving pseudoscalar TFFs and decay constants, which have been derived in hard-wall AdS/QCD models, and the \new{complete} hadronic light-by-light tensor. 

The transition form factor (TFF) of an individual pseudoscalar resonance is defined as
\begin{align}
    i \int d^4x e^{iq_1 x} \langle \Omega | J^{\mu}(x) J^\nu(0)| P_n(p)\rangle =\nonumber\\\varepsilon^{\mu \nu \alpha \beta}(q_1)_{\alpha} (q_2)_{\beta} F_n(Q_1^2,Q_2^2),
\end{align}
where $(q_1+q_2)^2=p^2=m_n^2$.
The function $F_n(Q_1^2,Q_2^2)$ determines the amplitude of the decay into two, in general, virtual photons and is a key ingredient in dispersive approaches to light-by-light scattering contributions to the muon $g-2$.

The TFF of the soft-wall model \new{derived from the Chern-Simons term} in the 
\new{isovector} $a=3$ sector is given by
\begin{align}
\label{eq:TFForiginalformula}
F_n(Q_1,Q_2)= \text{tr}(t^3 \Q^2  ) \frac{N_c}{2 \pi^2} \int_0^{\infty} \!\!dz \,\varphi_n'(z) \J(z,Q_1)\J(z,Q_2)
\end{align}
where 
\be
\J(z,Q)= \Gamma(\frac{Q^2}{4c^2}+1) U(\frac{Q^2}{4c^2},0,c^2z^2)
\ee
is the analytically known soft-wall vector bulk-to-boundary propagator, with $U(a,b,z)$ the Tricomi confluent hypergeometric function; $\Q$ is the quark charge matrix. For all $Q^2>0$, $\J$ goes from $1$ at $z=0$ to zero at $z= \infty$ strictly monotonically, and its large expansion $z$ for any $Q^2>0$ is given by
\begin{align}
\label{eq:Jbbzasympt}
    \J(z,Q) \sim {(c^2z^2)^{-\frac{Q^2}{4c^2}}} \Gamma(\frac{Q^2}{4c^2}+1).
\end{align}
Rewritten in terms of the mode function $y$ the TFF reads
\begin{align}
\label{eq:TFFlargen}
     F_n(Q_1^2,Q_2^2)&=\frac{N_c}{2 \pi^2}\text{tr}(t^3 \Q^2  ) g_5 \sigma \nonumber \\& \times \int dz e^{\frac{\phi}{2}} z^{5/2} y_n(z) \J(z,Q_1)\J(z,Q_2)
\end{align}
\new{in the chiral limit, or more generally at large mode numbers $n$.}

A crucial point is that when transforming $\partial_z \varphi$ to $y$, there is an exponential factor with a positive exponent from the dilaton,
\begin{align}
\label{eq:vartraf}
    \partial_z \varphi=  g_5 \sigma z^{5/2} e^{\frac{\phi}{2}} y.
\end{align}
Replacing the integration variable $z(u)= z_r \, u$
with $z_r$ given by \eqref{eq:zr}
we have
\begin{align}
      F_n & (Q_1^2,Q_2^2)=   \frac{N_c}{2 \pi^2}\text{tr}(t^3 \Q^2  ) g_5 \sigma z_r^{7/2} \nonumber \\ &\times \int_0^{\infty} du \;e^{\frac{c^2 z_r^2 u^2}{2}} u^{5/2} y_n(z_r u) \J(z_r u,Q_1)\J(z_r u,Q_2)
\end{align}
with the four regions defined above translated to $u$, where regions I and IV can be
safely neglected.
The contribution to the integral from region \textrm{II} is
\begin{align}
   F^{\mathrm{II}}_n(Q_1,Q_2)=    -\frac{N_c}{2 \pi^2}\text{tr}(t^3 \Q^2  )(g_5 \sigma)^{-3/4}\sqrt{\frac{2}{\pi}}C_1 E_n^{5/8} I(E_n)\
   \end{align}
   where
   \begin{align}
    &I(E_n)=\int_{u_1}^{u_2} u^{5/2}\exp \bigg\{ \frac{c^2}{2 g_5 \sigma}\sqrt{E_n} u^2 \bigg\}(1-u^4)^{-1/4}\nonumber \\ \label{eq:critintegral}
    &\times\cos \bigg( \frac{E_n^{3/4}}{\sqrt{g_5 \sigma}} \int_0^u \sqrt{1-u^{' 4}} du' \bigg)\J(z_r u,Q_1)\J(z_r u,Q_2).
\end{align}
The integrand is oscillating rapidly due to the cosine, and its average value rises rapidly with growing $u$, due to the exponential factor.
 Let us estimate the contribution $I^{VP}(E_n)$ of an arbitrary valley at $u=u_m$ together with its peak to the right at $$u=u_m+\Delta u=u_m+ \frac{1}{\sqrt{1-u_m^4}}\frac{\pi\sqrt{g_5 \sigma}}{E_n^{3/4}}.$$ 
 Since all the other terms in the integral are slowly varying on that scale, the result reads
 \begin{align}
 \label{eq:peakvalleestimate}
I^{VP}(E_n)\sim \frac{(\Delta u)^2}{(1-u_m^4)^{1/4}} \exp \bigg\{ \frac{c^2}{2 g_5 \sigma}\sqrt{E_n} u_m^2 \bigg\} \sqrt{E_n} u_m
\end{align}
times a slowly varying function.

At generic $u_m$, the expression above goes like $E_n^{-1} \exp \frac{c^2}{2 g_5 \sigma} \sqrt{E_n}$, which means that in the large $E_n$ limit, the peak is exponentially more important than the valley to the left of it. Similarly, any valley dominates the peak to its left.
In order to estimate the integral over region II we may then consider just the rightmost peak or valley. In the $u$ coordinate, the rightmost peak or valley of region II is at $u_c\sim 1-(E_n/c^2)^{-a_2}$, which upon insertion into \eqref{eq:peakvalleestimate} gives
\begin{align}
    I(E_n)\sim E_n^{-1+\frac{5}{4}a_2}\exp \left( \frac{c^2}{2 g_5 \sigma}\sqrt{E_n}  \right).
\end{align}
This depends on the precise way we choose to separate regions II and III since $a_2$ can take any value in $(0,\frac{1}{2})$. The bigger the value of $a_2$, the bigger the value of the integral gets. When increasing $a_2$, we at the same time shrink region III and describe more of the solution with the WKB approximation of region II.

This is really a hint that the contribution of region III will dominate that of region II. 
The integral should be dominated by the last peak or valley before the exponential decay of the solution in region \textrm{IV}.
This indeed lies in region \textrm{III} where the Airy function \eqref{eq:Airy III text} should be used. 
At large $E_n$ there will always be peaks that are included in region III, because the stationary points $z_i$ of the solution in region III are at $z^{\text{stat}}_i-z_r=-l_i (4 g_5^2\sigma^2)^{-1/3} z_r^{-1} \sim (E_n/c^2)^{-1/4}$ (with $l_i$ of order $1$), of which more and more are contained in $[z_2,z_3]$ as $E_n$ grows. Notice also that for the choice of $1/2>a_2>1/4$, the size of region III shrinks to zero (and gets pushed further to the right). Regardless of how $E_n$ grows, the oscillations grow faster and more and more oscillations will be contained in region III, whose last peak or valley dominates the integral for the TFF.

\begin{figure}[t]
    \centering
    \includegraphics[width=1\linewidth]{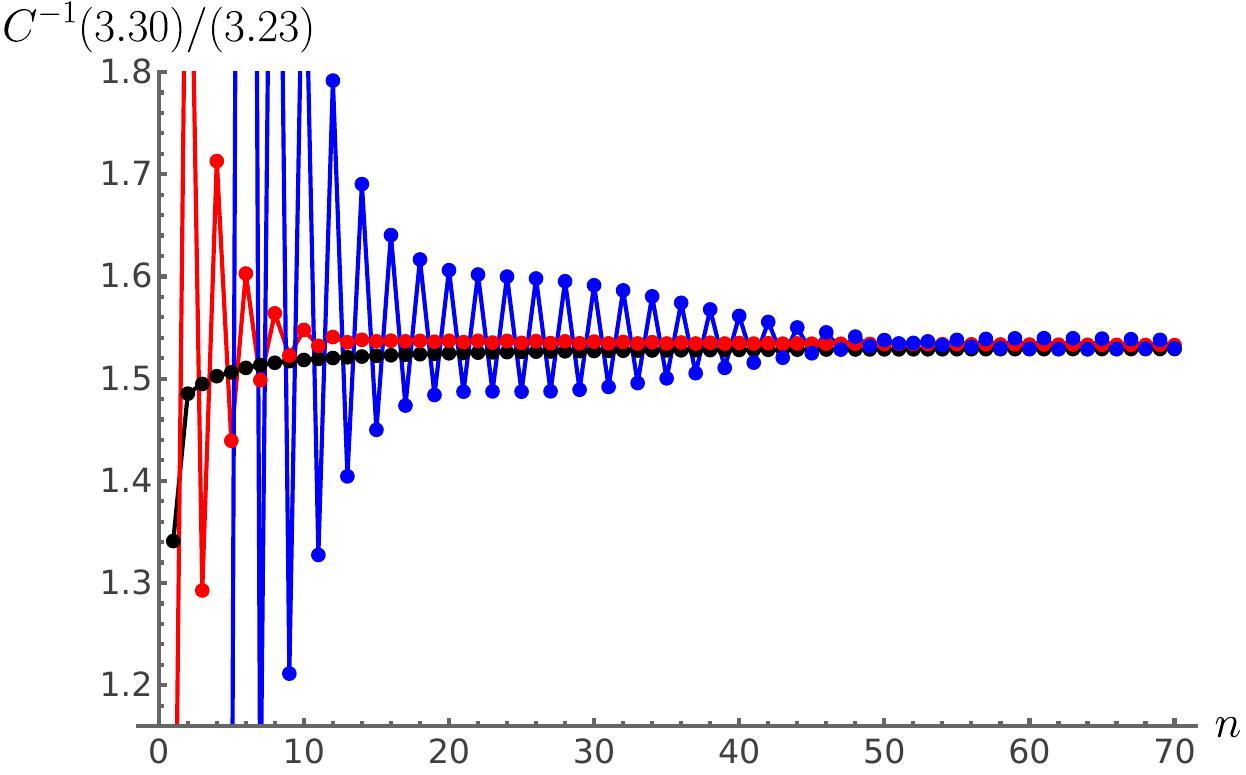}
    \caption{The ratio of \eqref{eq:TFFscaling} with $C$ omitted and \eqref{eq:TFFlargen} for $Q_i^2=0$ GeV$^2$ (black), $Q_i^2=1.5$ GeV$^2$ (red), and $Q_i^2=2.5$ GeV$^2$ (blue) as a function of the {mode number $n$} for the first 70 modes. The convergence to the analytic expression \eqref{eq:TFFscaling} is faster for smaller $Q_i^2$.}
    \label{fig:TFFratio}
\end{figure}

If we simply evaluate the integrand at the last peak of region III and multiply by the width of the peak, one obtains the following scaling behavior for the TFF \new{at large $E_n=m_n^2\sim n^{4/3}$} 
\begin{align}
\label{eq:TFFscaling}
    F_n(Q_1^2,Q_2^2) \sim & C (-1)^{n+1}  \; \sigma^{-\frac{1}{4}}\nonumber \\ \times & m_n^{-\frac{1}{4}} e^{\frac{c^2}{2 g_5 \sigma} m_n}  \J(z_r,Q_1)\J(z_r,Q_2)
\end{align}
with 
$z_r=\left(\frac{m_n}{g_5\sigma}\right)^{\frac{1}{2}}$ 
and $C$ a numerical constant of order 1 (independent of $E_n$ and $Q_i$). This is indeed larger than the contribution of region II for any choice of $a_2$.
We finally arrive at \eqref{eq:expscalsimple} 
by inserting the asymptotic expansion for the vector bulk-to-boundary propagator \eqref{eq:Jbbzasympt}, 
\be
\J(z_r,Q)\sim \Gamma(\frac{Q^2}{4c^2}+1) \left(\frac{c^2 m_n}{g_5 \sigma}\right)^{-Q^2/(4c^2)}
\ee
which is valid for $z_rQ\gg 1$.
In figure \ref{fig:TFFratio} we have displayed the ratio of the scaling \eqref{eq:TFFscaling} and the TFF \eqref{eq:TFFlargen} computed numerically within the WKB approximation for different $Q_i^2$, validating \eqref{eq:TFFscaling}.
We also see that convergence to the scaling law with constant $C$
sets in the later the larger $Q_i$ is.

Even though a contribution by an individual peak is completely suppressed by the contribution from the valley directly right of it, it is, in principle, conceivable that the sum over all such peaks and valleys can compete with the last peak that is situated in region III and whose contribution is estimated by \eqref{eq:TFFscaling}. For even mode numbers $n$, there is a peak for every valley, and the contribution from each such pair will have the same sign. A rigorous statement is then that the (absolute value of the) sum over all such contributions from the pairs is bounded \emph{from below} by the absolute value of \eqref{eq:TFFscaling}. 
For odd $n$ we can begin pairing up peaks and valleys starting from the rightmost peak in region III. Those pairs will again contribute with the same sign, while the leftmost peak will be unpaired and contributing with the opposite sign. The leftmost peak is located in region I at $u\sim E_n^{-1/4}$ for which the exponential factor is only of order one. The conclusion that the absolute value  \eqref{eq:TFFscaling} is a lower bound of the absolute value of the TFF \eqref{eq:TFForiginalformula} is therefore also true for odd $n$.

The exponential scaling with $E_n$ will be responsible for all sorts of convergence issues. Since we will be using the exponential scaling to show that certain expressions diverge, we will for \new{simplicity} act as if the exponential scaling were not just a lower bound (which was proven rigorously) on the TFF, but rather its true large $E_n$ behavior (for which we only have numerical evidence). 

Let us first investigate a holographic sum rule derived in \cite{Leutgeb:2021mpu} (and later generalized in \cite{Leutgeb:2022lqw} and \cite{Leutgeb:2024rfs}) that is relevant in the VVA correlator and which reads
\begin{align}
\label{eq:sumrule}
    \sum_n F_n(0,0)f_n =\frac{N_c}{2 \pi^2}\text{tr}(t^3 \Q^2  ).
\end{align}
From the exponential scaling of the TFF, it is clear that this cannot be obeyed in the present soft-wall model. The decay constant $f_n$ scales like $E_n^{-\frac{3}{8}}$ and goes to zero, but clearly the exponential behavior dominates this at some point and the sum on the LHS diverges.

Another important observable is the hadronic light-by-light scattering tensor $\Pi^{\mu \nu \lambda \rho}$. 
It is defined as
  \begin{align}
	\Pi^{\mu\nu\lambda\sigma}(q_1,q_2,q_3)&= -i \int d^4x \, d^4y \, d^4z \, e^{-i(q_1 \cdot x + q_2 \cdot y + q_3 \cdot z)} \notag\\
	&\hspace{-20pt}\times\langle 0 | T \{ j_\text{em}^\mu(x) j_\text{em}^\nu(y) j_\text{em}^\lambda(z) j_\text{em}^\sigma(0) \} | 0  \rangle .
\end{align}

It can be decomposed into in general $54$ singularity free tensor structures $T_i^{\mu \nu \lambda \rho}$:
\begin{align}
    \Pi^{\mu \nu \lambda \rho}(q_1,q_2,q_3)=\sum_i T_i^{\mu \nu \lambda \rho} \Pi_i.
\end{align}
A particular combination that enters into the anomalous magnetic moment is $\hat{\Pi}_1=\Pi_1+ (q_1 \cdot q_2) \Pi_{47}$. For the muon $g-2$, one can go to the simpler kinematics $q_4=q_1+q_2+q_3=0$. In terms of the TFFs of the tower of pseudoscalars, it reads
\begin{align}
    \hat{\Pi}_1= \sum_n \frac{F_n(-q_1^2,-q_2^2)F_n(-q_3^2,0)}{q_3^2-m_n^2}
\end{align}
Clearly, the WKB analysis of $F_n(Q_1^2,Q_2^2)$ implies that this sum does not converge for any $q_1^2,q_2^2,q_3^2$.

\begin{table}[t]
    \centering
    \begin{tabular}{c|c||c|c}
    $m_n$& $a^{\pi^{(n)}}_\mu\times 10^{11}$& $m_n$ & $a^{\pi^{(n)}}_\mu\times 10^{11}$ \\
    \hline
    0.135 & \new{75.36} & 8.037& 0.232 \\
        2.087 & 1.688 & 8.541& 0.237 \\
         3.030& 0.630& 9.031& 0.244     \\
         3.827& 0.426&  9.507&   0.254  \\
         4.544& 0.327&  9.972& 0.265   \\
         5.205& 0.281&  10.426&0.279   \\
         5.826& 0.254&  10.871& 0.295       \\
         6.414&0.240 &  11.306& 0.314    \\
         6.976&0.233 &   11.734 &  0.334   \\
         7.516&0.231 &  12.153 & 0.357  \\
    \end{tabular}
    \caption{The $a_\mu$ contributions of ground-state ($n=0$) and excited pions 
    ($n\ge1$) in the SW model
    up to $n=19$. 
    For $m_n$ above 8 GeV, their magnitude rises again, leading to a divergent
    sum $\sum_n a^{\pi^{(n)}}_\mu$.}
   \label{tab:g-2modesnew-mq}
\end{table}

\begin{table}[t]
    \centering
    \begin{tabular}{c|c||c|c}
    $\mA_n$ & $a^{A^{(n)}}_\mu\times 10^{11}[L+T]$ & $\mA_n$ & $a^{A^{(n)}}_\mu\times 10^{11}[L+T]$ \\
    \hline
        1.679 & 36.46 [20.39+16.07]& 9.289 & 0.392 [0.196+0.195]\\
        2.674 & $-$0.50 [$-$0.05$-$0.45]& 9.758 & 0.013 [0.010+0.002]\\
         3.503  & 4.05 [2.12+1.93]& 10.217&  0.349 [0.175+0.175]\\
         4.241  & $-$0.22 [$-$0.06$-$0.16]& 10.665& 0.030 [0.018+0.012]\\
          4.920 & 1.55 [0.80+0.75]&  11.105& 0.326  [0.163+0.164]\\
          5.554& $-$0.113 [$-$0.034$-$0.079]&  11.535& 0.047 [0.026+0.021]\\
           6.153& 0.87 [0.45+0.43]& 11.958 & 0.315 [0.157+0.158]\\
           6.725& $-$0.061 [$-$0.018$-$0.043]&  12.373& 0.065 [0.035+0.030] \\
            7.273& 0.60 [0.30+0.30]&12.782& 0.313 [0.156+0.157] \\
            7.801& $-$0.030 [$-$0.007$-$0.023]& 13.184 & 0.085 [0.044+0.040]\\
            8.312 & 0.466 [0.234+0.232]& 13.580 & 0.318 [0.158+0.160]\\
            8.807 & $-$0.007 [0.002$-$0.009]&13.970& 0.106 [0.055+0.051] \\
    \end{tabular}
    \caption{The $a_\mu$ contributions of the \new{ground-state ($n=1$)} axial vector multiplet in the SU(3)-symmetric case ($4\times a_\mu^{a_1}$ with uniform $m_q\neq0$) and its tower of excited states up to $n=24$. Odd and even numbered states contribute with different magnitudes,
    for $\mA$ below 9 GeV even with different sign, but above 13 GeV both grow with positive size, clearly indicating a divergent sum. \new{For two successive modes at $\mA\approx 20$ GeV with respectively odd and even $n$ we find $a_\mu\times10^{11}=0.685$ and $0.507$; a mode with $M^A \approx 30$ GeV already contributes $a_\mu \times 10^{11} \approx 8.7$.}}
   \label{tab:g-2modesnew-mq-axial}
\end{table}

Even without the use of the WKB approximation, one can see signs of this pathological behavior in the $g-2$ contribution by simply computing its value for 
\new{a sufficiently large number of}
excited states numerically. 
The results {(with $m_q$ {and $\sigma$ such that $m_{\pi^0}=134.9$ MeV and $f_\pi=93.1$ MeV to match \cite{Colangelo:2023een,Colangelo:2024xfh}})} are shown in Table \ref{tab:g-2modesnew-mq} where one can clearly see an increase in the $a_\mu$ contributions
\new{from excited pseudoscalars with masses above} $7.5$ GeV, \new{preventing a convergence of their sum}.

In the numerical evaluation, 
it is convenient to resort to a finite cutoff $z_0$, but this needs to be large enough,
otherwise the results will be misleading.\footnote{Our results for the individual mode contributions deviate slightly from
those obtained in \cite{Colangelo:2023een,Colangelo:2024xfh}. More importantly,
instead of the result given in \cite{Colangelo:2024xfh} for the sum over all axial-vector contributions, $41.3\times 10^{-11}$, we
obtain infinity as the cutoff is sent to infinity, also when using the full bulk-to-bulk propagator instead of
a sum over resonances, see Sec.\ \ref{sec:Lightbylighttensor}.}
One can either specify $\eta(z_0)=\varphi(z_0)$ corresponding to $\partial_z y(z_0)=0$, or $\partial_z \varphi(z_0)=0$ corresponding to $y(z_0)=0$ as boundary conditions at the cutoff.
If the cutoff is pushed to infinity, then both conditions will be satisfied and one is in the soft-wall regime.
If, however, the cutoff is always kept fixed at some $z=z_0$, then for masses $m_n>m_c$, where $m_c$ is some critical mass which depends on $z_0$, the second boundary condition will be more and more violated and one effectively \new{falls back to} a hard-wall model.

In order to avoid this, we perform our computations in terms of $y$ and adopt the following algorithm: We compute the normalized mode for some fixed cutoff $z_0$ with boundary condition $y(z_0)=0$ and we increase $z_0$ until $y'(z_0)$ drops below a fixed value.
For example (in units of GeV), for $y'(z_0)<10^{-6}$, 
the mode with $m \approx 3.8$ requires $z_0>5.74$, while the mode with $m\approx 8.5$ requires $z_0>6.67$.

In the axial-vector case, the amplitude
\begin{align}
 \mathcal{M}_A^{\mu \nu}=   i \int d^4x e^{iq_1 x} \langle \Omega | J^{\mu}(x) J^\nu(0)| A(p)\rangle =\mathcal{M}_A^{\mu \nu\alpha}\varepsilon_{\alpha}^A(p)
\end{align}
leads to 

\begin{align}
    \mathcal{M}_A^{\mu \nu \beta}&=-i \frac{N_c}{4 \pi^2} \varepsilon^{\mu \tilde{\nu} \beta \alpha} (q_1)_{\alpha}(\delta^{\nu}_{\tilde{\nu}}q_2^2-(q_2)^{\nu} (q_2)_{\tilde{\nu}})A(Q_2^2, Q_1^2) \nonumber \\ &+(\text{crossed term})
\end{align}
with
\begin{align}
\label{eq:axtff}
    A_n(Q_1^2,Q_2^2)=\frac{2}{Q_1^2}\text{tr}(t^3 \Q^2  ) \int dz A^{\perp}_n(z) \J'(z,Q_1) \J(z,Q_2)
\end{align}
holographically.

The analysis for large $M_n^A$ is very similar to the pseudoscalar case one, so we will not repeat it in detail here. Going through the steps as in the pseudoscalar case, we find the scaling
\begin{align}\label{eq:ATFFscaling}
     A_n(Q_1^2,Q_2^2)\sim \frac{C'}{Q_1^2}\J'(z_r,Q_1) \J(z_r,Q_2)  e^{\frac{c^2}{2g_5 \sigma} \sqrt{E_n}} E_n^{-\frac{1}{8}} \sigma^{-\frac{1}{4}}
\end{align}
\new{with $C'$ a} numerical constant independent of $E_n,Q_i$.
Therefore also the axial-vector contribution to the muon $g-2$ will diverge. 

\begin{figure}[t]
    \centering
    \includegraphics[width=\linewidth]{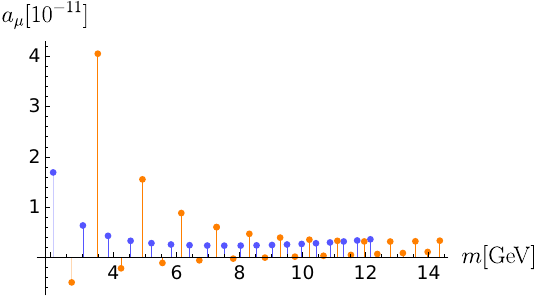}
    \caption{Plot of the SW results for the $a_\mu$ contributions of excited pseudoscalars and excited axials, Table \ref{tab:g-2modesnew-mq} (blue) and \ref{tab:g-2modesnew-mq-axial} (orange)}
    \label{fig:amuPS-SW}
\end{figure}

Table \ref{tab:g-2modesnew-mq-axial} lists the full numerical 
$a_\mu$ contributions of axial-vector
mesons and their excited states, with $a_\mu^A=4 a_\mu^{a_1}$ corresponding to
the SU(3)-symmetric results for $a_1$, $f_1$, and $f_1'$ combined.
Here, even and odd mode numbers differ in magnitude and, for small $n$, also in sign.\footnote{This is rather different from the various HW models considered in \cite{Leutgeb:2021mpu}. There, the contributions were all positive and monotonically decreasing with mode number.}
However, both increase eventually with a positive sign, making it impossible
to cancel the divergent contribution of the pseudoscalars, but only adding to the latter.
(The $a_\mu$ contributions of pseudoscalars and axials
are shown together in Fig.~\ref{fig:amuPS-SW}.)

We emphasize
that the exponential scaling 
found in \eqref{eq:TFFscaling} and \eqref{eq:ATFFscaling}
is not restricted to only the transition form factors. Any coupling between 4 dimensional degrees of freedom derived from the Chern-Simons term in the soft-wall model will have this property.

The WKB machinery may also be used for hard-wall models to compute transition form factors for large $n$. We give closed formulas for $A_n(Q^2,Q^2)$ and $A_n(Q^2,0)$ in \ref{app:HWTFF} and comment on the realization of the Melnikov-Vainshtein constraint. The asymptotic scaling laws of \cite{Hoferichter:2020lap} are recovered for $Q \gg \sqrt{E_n}=M_n^A$, but in this limit they do not contribute to the short-distance constraints anymore.

\section{VVA correlator and non-commutativity of limits}
\label{sec:noncomm}

The subject of this section is the VVA correlator
\new{in the SW model \cite{Colangelo:2011xk}}, 
which will be analyzed with WKB techniques. 
Similarly to the HLbL tensor, we show that when the VVA correlator is written as a sum over intermediate particle contributions, one encounters a divergent expression.
\new{However, when this observable is evaluated without resorting to a decomposition in terms of resonances, a well-defined finite result is obtained.}
The sole effect of the exponential scaling of the TFF on this observable is then that one cannot write it as a sum of contributions of individual intermediate particles, which is somewhat at odds with large-$N_c$ intuitions. The VVA correlator in the soft-wall model is still well defined, and this non-decomposability can be traced back to a non-commutativity of certain limits. 

This section also serves as a warm-up for section \ref{sec:Lightbylighttensor} (although being no 
prerequisite) in which we will similarly compute the HLbL tensor without resorting to a decomposition in terms of modes. Contrary to what happens in the current section, the HLbL tensor will be shown to diverge also in that case, which 
\new{can be avoided only by a modification of the model itself}.

The VVA correlator $\mathcal{W}_{\mu \nu \rho}(q_1,q_2)$ is defined by
\begin{align}
    \mathcal{W}_{\mu \nu \rho}(q_1,q_2)=i \int d^4x {d^4y} \, e^{i(q_1x+q_2y)}\langle J^{\text{em}}_\mu(x) J^{\text{em}}_\nu(y) J^{5}_\rho(0) \rangle,
\end{align}
where $J^5_{\mu}=\bar{\psi} \gamma_{\mu} \gamma_5 \Q_5 \psi$ is the axial current.
The above 3-point function can be computed holographically from the Chern-Simons term. The effective action reads
\begin{align}
S&=S_{\text{bulk}}+S_{\text{bdry}}\\
  S_{\text{bulk}}&=  \frac{N_c}{4 \pi^2} \text{tr} \int_{AdS_5} A (dV)^2 \\
  S_{\text{bdry}}&= -\frac{N_c}{12 \pi^2} \int_{\partial AdS_5} A(dV V+VdV) 
\end{align}
As explained in \cite{Cappiello:2019hwh}, the resulting 3-point function describes the consistent anomaly; here, however, we want to calculate a correlation function of covariant currents. Therefore, a Bardeen-Zumino counterterm has to be added \cite{Bardeen:1984pm}, which has the effect of canceling the contribution from $S_{\text{bdry}}$.

We will be content to work in the chiral limit in this section, which already exhibits all the phenomena we want to showcase.
In the soft-wall model, one finds
\begin{align}
    \mathcal{W}_{\mu \nu \rho}&=-\frac{N_c}{2 \pi^2} \text{tr} \left( \Q^2 \Q_5 \right)P_{\mu}^{\hat{\mu}}(q_1)P_{\nu}^{\hat{\nu}}(q_2)\varepsilon_{\hat{\mu} \hat{\nu} \lambda \alpha} \\ & \times\int_0^\infty dz\left(\mathcal{A}^{\perp}(z,q) P^\lambda_{\;\;\rho}(q) + \varphi(z,q) \frac{q^\lambda q_\rho}{q^2} \right) \nonumber \\ &\times\bigg(q_2^{\alpha} \J'(z,Q_1){\J(z,Q_2)}-q_1^{\alpha} \J'(z,Q_2)\J(z,Q_1) \bigg).
\end{align}
 $\mathcal{A}^{\perp}(z,q)$ solves the axial equations of motion \eqref{eq:axeq} with $(M_n^A)^2$ replaced by $q^2$, subject to the boundary condition $\mathcal{A}^{\perp}(0,q)=1$. Similarly $\varphi(z,q)$ together with $\eta(z,q)$ solve the pseudoscalar equations \eqref{eq:pseq} (with $m_n^2 \rightarrow q^2$) obeying $\varphi(0,q)=1$, $\eta(0,q)=0$.

We will be mainly interested in the $q_2\rightarrow 0$ limit which is relevant in the muon $g-2$. As detailed in \cite{Ludtke:2024ase} the amplitude can then be decomposed into two tensor structures $\tau_i$ and structure functions $\bar{\mathcal{W}}_i$
\begin{align}
      \mathcal{W}_{\mu \nu \rho}(q_1,q_2)= \sum_{i=1}^2 \bar{\mathcal{W}}_i(q_1^2) q_2^{\sigma} (\tau_i)_{\mu \nu \rho \sigma}(q_1)+ \mathcal{O}(q_2^2).
\end{align}
Projecting the holographic expression on these tensor structures yields
\begin{align}
    \bar{\mathcal{W}}_1(q^2)&=-\frac{N_c}{2 \pi^2} \text{tr}  \left(\Q^2 \Q_5 \right) \int_0^{\infty} dz \; \frac{\varphi(z,q)}{q^2} \J'(z,Q) \\
    \label{eq:omegaT}
    \bar{\mathcal{W}}_2(q^2)&=\frac{N_c}{2 \pi^2} \text{tr}  \left(\Q^2 \Q_5 \right) \int_0^{\infty} dz \;\frac{\mathcal{A}^\perp(z,q)}{q^2} \J'(z,Q)
\end{align}

In the chiral limit $\varphi(z,q)=1$ for all $q^2$; hence, $\bar{\mathcal{W}}_1=\frac{N_c}{2 \pi^2} \text{tr}  \left(\Q^2 \Q_5 \right) \frac{1}{q^2}$.

There is no such simplification for $\bar{\mathcal{W}}_2$, but at large Euclidean $Q^2=-q^2$ we can apply the WKB approximation to solve approximately for $\mathcal{A}^\perp$.
The boundary condition $\mathcal{A}^{\perp}(0,Q)=1$ translates to $\xi=1/\sqrt{z}$ when transforming to the Schrödinger form $\mathcal{A}^{\perp}= \frac{\xi}{\sqrt{\rho}}$ with $\rho=\frac{e^{-\phi}}{z}$.
Then with $\kappa=\sqrt{V_A+Q^2}$ the condition $\frac{\kappa'}{\kappa^2} \ll 1$ is obeyed everywhere except at very small $z$, where we use the analytic solution
\begin{align}
\label{eq:bulktoboundaryaxial}
    \xi(z)=Q \sqrt{z} K_1(Qz),
\end{align}
\new{which is valid for $z\ll \mathrm{min}(c^{-1},(g_5\sigma)^{-1/3})$ and}
which goes like $\frac{1}{\sqrt{z}}$ at small $z$.
The other linearly independent solution $\sqrt{z}I_1(Qz)$ could be added without violating the boundary condition at $z=0$, but it goes like $e^{Q z}$ for large $Qz$, which cannot be matched to the WKB solution.

Matching the small-$z$ solution \eqref{eq:bulktoboundaryaxial}
to the WKB expression
\begin{align}
    \xi(z)=\frac{C_A}{\sqrt{\kappa}}\exp\left[-\int_{z_A}^z dz'\,    \kappa(z') \right]
\end{align}
at $z_A=Q^{-1} (Q/c)^{\alpha}$ with $0<\alpha<1$ gives $C_A=e^{-Q z_A} \sqrt{\frac{\pi}{2}}Q$.

The point $z_A$ scales with $Q$ such that $z_A \rightarrow 0$ and $z_A^2 Q^2 \rightarrow \infty$ when $Q \rightarrow \infty$. As in the discussion of the $z_i$ of the previous section, $z_A$ is designed \new{in analogy to $z_1$} such that both the small-$z$ solution and the WKB solution are valid for large $Q$.

The small-$z$ region of the integral \eqref{eq:omegaT} is finite because the approximate solution correctly implements the boundary condition and $\partial_z \J \sim z$. At large $z$, 
\be
\mathcal{A}^{\perp}=e^{\frac{c^2 z^2}{2}} \sqrt{z} \, \xi(z) \sim e^{-g_5 \sigma \frac{z^3}{3}},
\ee
so we expect the integral \eqref{eq:omegaT} to converge and give a finite value. 

Although one cannot use the WKB approximation to calculate $\bar{\mathcal{W}}_2$ at small $Q$ reliably, one can still use it to show that the integral at least converges for all $Q$ since $\frac{\kappa'}{\kappa^2}\sim z^{-3}$ at large $z$ for arbitrary $Q$ which can be made as small as one likes by going to large $z$.
The criterion of the validity of the WKB approximation is therefore obeyed at large $z$ for arbitrary $Q$.

Since the solution \eqref{eq:bulktoboundaryaxial} is exact at small $z$ and the solution in-between is smooth and bounded, \eqref{eq:omegaT} converges for all $Q$.

On the other hand, one may calculate the pole contribution of an individual axial vector meson to \eqref{eq:omegaT} from standard QFT polology, which reads
\begin{align}
\label{eq:polecontr}
      \bar{\mathcal{W}}_2^\mathrm{pole}(Q^2)&=-\frac{N_c}{2 \pi^2} \text{tr}  \left(\Q^2 \Q_5 \right)   \frac{F_A}{Q^2+(M^A)^2}\frac{A(Q^2,0)}{2} \nonumber \\
     &=-\frac{N_c}{2 \pi^2} \text{tr}  \left(\Q^2 \Q_5 \right)   \frac{F_A}{Q^2+(M^A)^2} \nonumber\\ &\times \int dz \,A^{\perp}(z) \frac{\partial_z \J (z,Q)}{Q^2}
\end{align}
with $F^A$ being an axial-vector decay constant and $A(Q^2,0)$ the singly virtual axial TFF.
At large $N_c$, the only singularities one would expect in correlation functions and amplitudes (at leading order) are poles, and a dispersive analysis would reconstruct the VVA correlator simply as a sum over all such pole contributions \eqref{eq:polecontr} (except for potential subtractions).
According to the discussion in the previous section, the TFF $A_n(Q^2,0)$ scales exponentially, $\sim e^{\frac{c^2}{2 g_5 \sigma}\sqrt
E_n}$, and the sum over all pole contributions does not converge. This is \emph{seemingly} in contradiction to the above analysis, where it was concluded that $\bar{\mathcal{W}}_2$ is finite when calculated using the axial bulk-to-boundary propagator.
In Appendix \ref{app:nonuniform}, we trace this back to a lack of
uniform convergence of the mode expansion, which does not permit
to swap the infinite sum over modes with the indefinite integration
over the holographic coordinate $z$.

The divergences investigated in this section seem to be mostly harmless. The relevant 3-point functions are well defined although it is not possible to decompose them into a sum over modes, which is slightly unnatural from a large-$N_c$ perspective. In the next section, we will encounter more severe divergences that cannot be avoided without modifying the model itself.

\section{Hadronic light-by-light tensor from the 5-dimensional approach}
\label{sec:Lightbylighttensor}

In this section, we investigate the hadronic light-by-light tensor from the 5-dimensional point of view. Above we saw that for the VVA correlator the sum over pole contributions did not give a convergent result, but in the process of arriving there, integrals and sums were illegally exchanged, thus invalidating the procedure. Now we try to answer the question of whether something similar happens in the HLbL tensor.

\begin{figure}[h]
    \centering
    \includegraphics[width=1\linewidth]{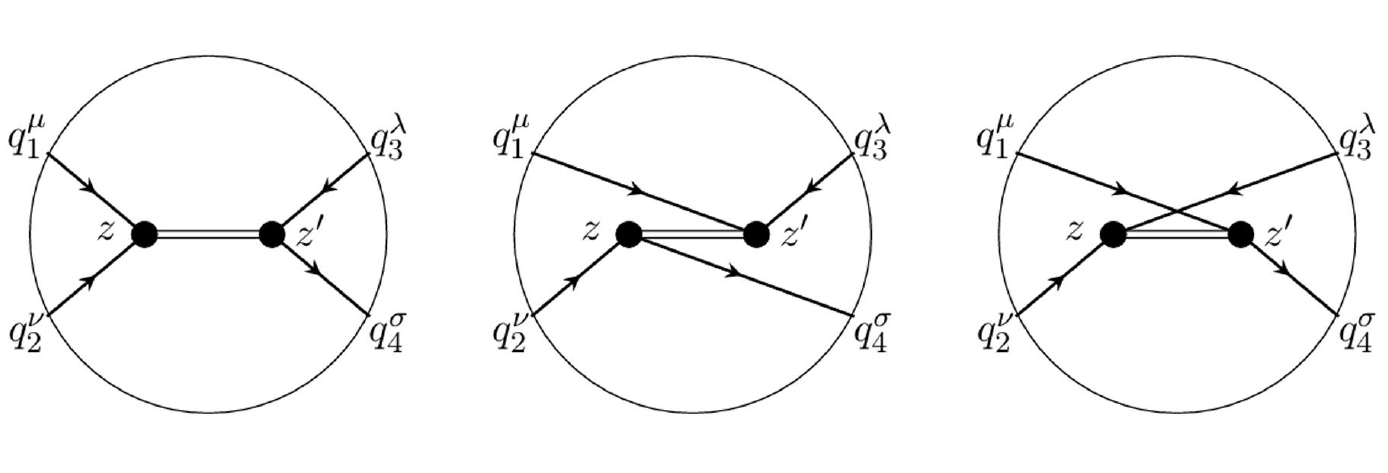}
    \caption{The Witten diagrams needed in the calculation of the HLbL tensor. Single lines denote a vector bulk-to-boundary propagator, while double lines denote a bulk-to-bulk propagator of the axial gauge field $A_{\mu}(x,z)$.}
    \label{fig:Wittdiagr}
\end{figure}

To leading order\footnote{At subleading order in $N_c$ there are contributions with four AAV vertices connected by an axial loop in the bulk. Those would describe pion loops in 4d among other things.} in $N_c$ the HLbL tensor is calculated by the Witten diagrams in Fig.\ \ref{fig:Wittdiagr}.

The resulting tensor is a highly complicated object which we now show to be actually divergent in the soft-wall model, \new{i.e., also when the full bulk-to-bulk propagator
is employed}.

The bulk-to-bulk propagator of the axial gauge field in $A_z=0$ gauge can be decomposed as
\begin{align}
    G^{\mu}_{\; \; \nu}(z,z',q)=G_A(z,z',q^2)P^{\mu}_{\perp\;\; \nu}(q)+ G_{\varphi \varphi}(z,z',q^2)q^{\mu}q_{\nu}
\end{align}
with $P^{\mu}_{\perp\;\; \nu}(q)=\delta^{\mu}_{\nu} -\frac{q^{\mu}q_{\nu}}{q^2}$.
The (as we call them) pseudoscalar and axial-vector bulk-to-bulk propagators $(G_{\varphi \varphi},G_{\eta \varphi}),G_A $ obey the following equations
\begin{align}
&\frac{q^2}{g_5^2} \partial_z(\frac{e^{-\phi}}{z} \partial_z G_{\varphi \varphi})+q^2\frac{v^2}{z^3}e^{-\phi}(G_{\eta \varphi}-G_{\varphi \varphi})= \delta(z-z') , \\
  &  \partial_z (\frac{v^2 e^{-\phi}}{z^3} \partial_z G_{\eta \varphi})+ \frac{q^2 v^2}{z^3}e^{-\phi}(G_{\eta \varphi}-G_{\varphi \varphi})=0,\\
  \label{eq:Axgreenf}
&    \frac{1}{g_5^2} \partial_z(\frac{e^{-\phi}}{z} \partial_z G_{A}) + \frac{q^2 e^{-\phi}}{g_5^2 z } G_A-\frac{v^2 e^{-\varphi}}{z^3} G_A=\delta(z-z') ,
\end{align}
subject to normalizable boundary conditions. As above, we consider only the
flavor-symmetric case, focusing on the chiral limit without U(1)$_A$ anomaly.

The axial-vector contribution (in the $g-2$ kinematics $q_4=0$) to $\hat{\Pi}_1$ reads
\begin{align}
    \hat{\Pi}_1=-g_5^2\left(\frac{N_c}{12 \pi^2}\right)^2  \iint dz  dz' \; G_A(z,z',0) \Jpq(z,Q_3) \nonumber \\ \times\Big(q_1^2 \Jpq(z',Q_1)\J(z',Q_2)+q_2^2 \Jpq(z',Q_2) \J(z',Q_1)\Big)
\end{align}
where $\Jpq(z,Q)= \frac{\J'(z,Q)}{q^2}$. The crucial point will revolve around the large $z,z'$ integration region. There we may use the asymptotic expansions of $\J, \Jpq$ given in \eqref{eq:Jbbzasympt}. Actually, the only detail that we need is that for fixed momenta, $\J, \Jpq$ are well approximated by powers of $z,z'$.
The axial contribution to the large $z,z'$ integration region of \emph{any} of the $\hat{\Pi}_i$ is then of the form
\begin{align}
\label{eq:HLbL schematic}
    \hat\Pi= \iint dz dz'\; G_A(z,z',q) p(z,z')
\end{align}
with $p(z,z')$ obeying a power-law decay at large $z,z'$ and $q$ being some intermediate momentum. The function $p(z,z')$ also depends on momenta, although this is irrelevant here.

Rigorous statements about the $z,z'$ behavior of $G_A$ and the convergence of the integral in \eqref{eq:HLbL schematic} can be made most easily within the WKB approximation at large Euclidean $q^2=-Q^2$.
The Green's function is symmetric $G_A(z,z',Q)=G_A(z',z,Q)$ and in the Schrödinger form $G_A=\frac{1}{\sqrt{\rho}} \tilde{G}$ equation \eqref{eq:Axgreenf} 
reads
\begin{align}
\label{eq:btbschrödeq}
    -\tilde{G}''+(Q^2+V_A)\tilde{G}=-\frac{g_5^2}{\sqrt{\rho}} \delta(z-z') 
\end{align}
with $\rho= \frac{e^{-\phi}}{z}$ and $V_A=\frac{3}{4z^2}+c^4 z^2+g_5^2 \sigma^2 z^4$.
Integrating the equation with $\int_{z'-\varepsilon}^{z'+ \varepsilon} dz$ gives
\begin{align}
\label{eq:matchcond}
    \tilde{G}'(z'+\varepsilon)-   \tilde{G}'(z'-\varepsilon){=\frac{g_5^2}{\sqrt{\rho(z')}}  =g_5^2 e^{\frac{\phi(z')}{2}} \sqrt{z'}}.
\end{align}
The homogeneous version of \eqref{eq:btbschrödeq} has two linearly independent solutions $\xi_L(z), \, \xi_R(z)$, where we take $\xi_L \sim z^{3/2}$ near $z=0$ and $\xi_R \rightarrow 0$ for large $z$.
Then one can solve the matching conditions to get 
\begin{align}
\label{eq:bulktoboundaryaxialfin}
    G_A(z,z')&=\frac{g_5^2}{\sqrt{\rho(z)} \sqrt{\rho(z')}} \nonumber\\
    &\times\frac{ \theta(z'-z)\xi_R(z') \xi_L(z)+\theta(z-z')\xi_L(z') \xi_R(z)}{\xi_L'(z')\xi_R(z')- \xi_L(z') \xi_R'(z')} .
\end{align}
Noting that the Wronskian $\xi_L'(z')\xi_R(z')- \xi_L(z') \xi_R'(z')$ is independent of $z'$, the symmetry $G_A(z,z',Q)=G_A(z',z,Q)$ is apparent.
Explicit formulas for $\xi_L, \, \xi_R$ in the WKB approach are given in 
Appendix \ref{app: Btbb}.

The bulk-to-bulk propagator obeys $\lim_{z\rightarrow \infty}G(z,\kappa z)=0$ for all $\kappa \neq 1$, and also vanishes in the limit where one point is held fixed and the other goes to infinity. It is also well behaved near $z=0$ (and by symmetry near $z'=0$). 

\begin{figure}
    \centering
    \includegraphics[width=1\linewidth]{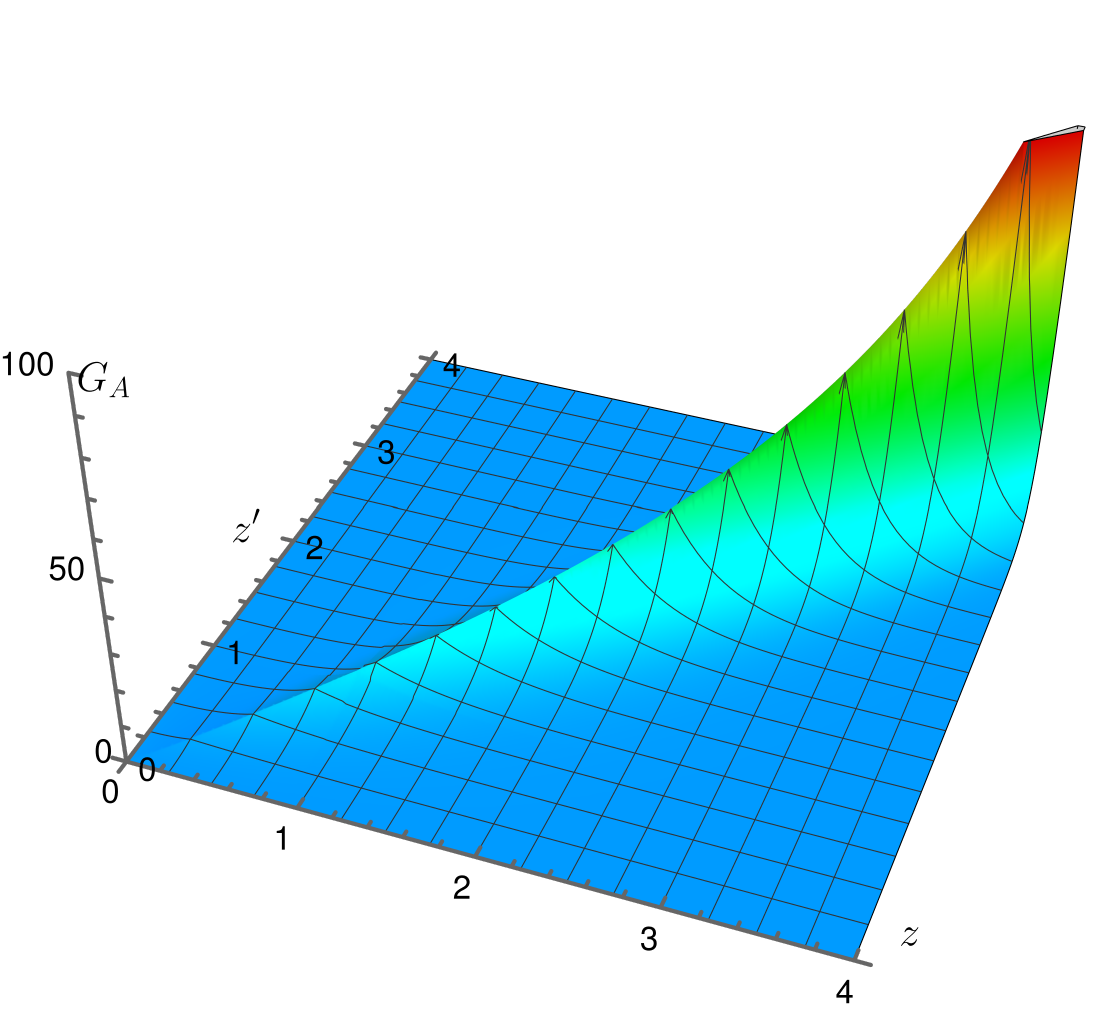}
    \caption{The axial bulk-to-bulk propagator $G_A(z,z',Q)$ for $Q=3$ in units of GeV {as a function of $z$ and $z'$ (in units of GeV$^{-1}$)}. The clearly discernible exponential rise along the diagonal is responsible for the divergence of $z,z'$ integrals in the computation of $\hat{\Pi}_i$.}
    \label{fig:bulktobulk}
\end{figure}

The other parts of the integrand, namely $p(z,z')$, also vanish in these limits sufficiently fast for generic $Q_i$, so convergence of the integral seems possible. There is, however, one problematic region along the diagonal $z=z'$, where $G \sim e^{c^2 z^2}$ (times power-law factors).
It is this exponential increase along the diagonal of the bulk-to-bulk propagator that will cause problems in the HLbL tensor.
A plot of the bulk-to-bulk propagator is shown in figure \ref{fig:bulktobulk}.
We focus on the subregion of points $z,\, z'$ whose distance to the diagonal $z=z'$ is less than $\frac{\delta}{2}$, which is taken to be fixed but much smaller than any length scale in the theory. Recalling that ${\kappa'}/{\kappa^2} \sim {z^3}/{(Q^2+g_5^2 \sigma^2 z^4)^{3/2}}$ at large $z$, we see that even for small momentum $Q$ the WKB approximation holds for large enough $z,z'$. In addition to $z,z'$ being close to the diagonal, we also assume that they are large enough for $\frac{\kappa'}{\kappa^2} \ll1$ to hold.

By symmetry, we only have to integrate below the diagonal\footnote{The other $z,\,z'$ dependent factors from the vector bulk-to-boundary propagators are not symmetric in $z, \, z'$ for generic $Q_i$, but inside the integral only the symmetric part contributes.}. 
Below the diagonal at large $z,z'$, the WKB approximation detailed in Appendix \ref{app: Btbb} predicts the axial bulk to bulk propagator to scale like
\begin{align}
   G_A(z,z',Q)\sim \frac{1}{\sqrt{zz'}}e^{\frac{c^2}{2}(z^2+z'^2)} e^{\frac{g_5 \sigma}{3}(z'^3-z^3)}
\end{align}

Using this expression, the integral has the structure
\begin{align}
\hat\Pi\sim    \int dz \int dz' \, \tilde{p}(z,z') e^{\frac{c^2(z^2+z'^2)}{z}} e^{\frac{g_5 \sigma}{3} (z'^3-z^3)}
\end{align}
where $\tilde{p}(z,z')$ collects all the terms with power-law behavior. 
We switch to variables $y=\frac{z+z'}{2}, \, \eta = \frac{z-z'}{2}$ where the integral at large $y$ becomes
\begin{align}
  & \int dy e^{c^2 y^2} \int_0^{\frac{\delta}{2}} d\eta \, e^{-2g_5 \sigma y^2 \eta+c^2 \eta^2-\frac{2g_5 \sigma}{3} \eta^3} \tilde{p}(y+\eta, y- \eta) \\
   &   \sim \int dy e^{c^2 y^2} \int_0^{\frac{\delta}{2}} d\eta \, e^{-2g_5 \sigma y^2 \eta} \tilde{p}(y,y) .
\end{align}
With $\delta$ being small, we can neglect everything but the term linear in $\eta$ in the exponent.
After performing the integral over $\eta$ one gets for large $y$
\begin{align}
    \hat{\Pi}   &\sim -\int dy\, e^{c^2 y^2} \frac{\tilde{p}(y,y)}{2 g_5 \sigma y^2},
\end{align}
which diverges for any $\tilde{p}(y,y)$ that has power-law behavior. Therefore the integral over a thin strip of width $\delta$ around the diagonal is infinite. As our derivation was insensitive to the detailed form of $p(z,z')$, all $\hat{\Pi}_i$ have this type of divergence. 
The integral is well behaved in all other regions and cannot undo the divergence coming from the thin strip.

A numerical manifestation of this divergence is presented in table \ref{tab:nucutoff} where we compute a cutoff version of $\hat\Pi_1$. The upper integration boundary $b_2$ is taken to larger and larger values while the lower integration boundary is fixed at $b_1=10$ GeV$^{-1}$.

\begin{table}[t]
    \centering
    \begin{tabular}{c|cccccc}
       $b_2$  &11 &12&13&14 &15 & 16 \\
       \hline
       $-\hat{\Pi}_1 \times 10^{2}$ & 0.06&0.63 &7.9&143 &3859 &1.487 $\times 10^5$
    \end{tabular}
    \caption{Cutoff dependence of $\hat{\Pi}_1$ for $Q_i=1$. As discussed in the text, this goes to infinity as $b_2 \rightarrow \infty$. This is true for all $Q_1,Q_2,Q_3$. The units of $b_2$ and $\hat{\Pi}_1$ are GeV$^{-1}$ and GeV$^{-4}$ respectively. A small strip of width $0.1$ dominates the integral.}
    \label{tab:nucutoff}
\end{table}

As the divergence originates from the large-$z$ region and seems to be a very complicated function of the momenta, it cannot simply be subtracted using holographic renormalization.
The pseudoscalars also have a very similar divergence, and it cannot cancel the divergences from the axial sector, since the pseudoscalars only contribute to $\hat{\Pi}_1$, while the axials contribute to more structure functions.

Therefore  also the 5-dimensional approach to the HLbL tensor gives a divergent result. This is in contrast to the situation that we encountered for the VVA correlator. There at least the 5-dimensional calculation using the axial bulk-to-boundary propagator gave a finite result, even though the mode sum was divergent. We interpret this as a breakdown of the holographic soft-wall model and conclude that it is not suitable for describing HLbL scattering and the $g-2$ of the muon.

\section{Variations of the soft-wall model} 
\label{sec:varsandkiritsis}

In this section, we examine some ways to improve the standard soft-wall model 
considered 
in the previous sections and investigate whether the divergent mode sums may be avoided.

We consider first modifications of the scalar, metric, and dilaton backgrounds, and then also modifications of the Chern-Simons term. We show that upon demanding Regge behavior for the vector-meson sector and a vacuum preserving the vector flavor symmetry, there seem to be only a handful of ways to avoid the exponential scaling in the TFF. Either $w^2(z)v^2(z)$ diverges at some finite $z=z_0$ (where $w(z)$ is the warp factor), which renders the spectrum hard-wall like, or it diverges at $z\rightarrow \infty$ in an exponential way. A third possible option is to replace the original Chern-Simons term by a scalar-extended version, where the gauge fields and the bifundamental scalar are grouped together into a superconnection.

In the final subsection, we will examine V-QCD for which the latter option is realized.

\subsection{Background modifications}

One simple modification of the above model is to demand that the background field $X_0$ solves the equations of motion and
\new{is bounded} on all of AdS$_5$ \new{so that backreactions of this background to the metric and the dilaton can be neglected in the 't Hooft limit of large $N_c$ with small $N_f$}. We may assume that a general potential $U(X X^\dagger)$ has been included in the action, which controls the precise behavior of the background solution \new{as in the improved hQCD models of \cite{Kwee:2007nq,Ballon-Bayona:2021ibm}, where a finite limit $X_0 \rightarrow \text{const.}$ as $z\rightarrow \infty$ has been advocated for}. 
We will now show that in this seemingly self-consistent scheme, not even individual TFFs are well defined.

Since $X_0$ goes to a constant, the Schrödinger potential \new{for pseudoscalars} now goes like $V_P= c^4z^2$ for large $z$, \new{which yields asymptotically linear Regge trajectories}, and 
\begin{align}
    y_n\sim z^{\frac{E_n}{2c^2}-\frac{1}{2}}e^{-\frac{c^2z^2}{2}}.
\end{align}
However, this implies that $\varphi_n'(z)$ in \eqref{eq:TFForiginalformula}
goes like
\begin{align}
    \varphi'_n(z)=g_5 v z^{-1/2}e^{\phi/2}y_n(z)\sim z^{\frac{E_n}{2c^2}-1},
\end{align}
while $\mathcal{J}\sim z^{-\frac{Q^2}{2c^2}}$,
rendering the integral defining $F_n(Q_1^2,Q_2^2)$ divergent for all
$Q_1^2+Q_2^2 \le E_n$.

Therefore, with a dilaton profile that goes like $e^{-c^2 z^2}$ asymptotically, a scalar background $X_0$ that goes to a constant leads to even bigger problems than before. 

One may also start from completely general warp factor $w(z)\neq 1/z$, dilaton $e^{-\phi}$, and scalar background $X_0$, \new{without any restrictions regarding their large $z$ behavior.}
\new{We will not assume any specific form of the equations of motion or boundary conditions that govern the background. These are, in general, difficult to solve and depend on which terms one decides to include in the action. We will only assume that the action \emph{contains} the kinetic term of the gauge fields and the bifundamental scalar
\begin{align}
\label{eq:actionassump}
   S \supset \int \sqrt{g} e^{-\phi}\,\text{tr}(|F_L^2|+|F_R|^2) +\int e^{-\phi}\sqrt{g} |DX|^2
\end{align}
 and that $X_0$ is diagonal in the flavor indices. This implies that the mass spectrum of vector mesons (at least for $a=0,3,8$) is solely determined by the kinetic term of the gauge fields in terms of $w(z), e^{-\phi}$. We will additionally demand that the background solution is such that asymptotically the vector meson trajectory is Regge-like. This demand will allow us to fix a combination of $w(z), e^{-\phi}$ without solving for the background equations of motion.
 This approach is therefore rather general since it does not assume any specific action for the dilaton, the metric or a potential $\tilde{U}(XX^\dagger,\phi)$. Regge behavior of vector mesons is only a rather minimal demand, and is something any soft-wall model should aim to reproduce anyway. 
 }

The potential for the vector mesons is determined by $\rho(z)=w(z)e^{-\phi}$
\begin{align}
    V_V=\frac{1}{2}\frac{\rho''}{\rho}-\frac{1}{4}\left(\frac{\rho'}{\rho}\right)^2.
\end{align}

Demanding asymptotically linear Regge trajectories for the vector mesons means $V_V \sim z^2$ at large $z$. In order to have a potential that is bounded from below, the first term in the above expression by itself needs to go like\footnote{If we take the first term to scale like $z^\alpha$, with $\alpha>2$, then the whole $V_V\sim z^\alpha$, as the second term cannot cancel the $z^\alpha$ behavior from the first term completely.} $z^2$.

Solutions to $\rho''(z)= \left(4 c^4 z^2 +r(z)\right)\rho(z)$ for large $z$, where $r(z)$ are \new{subdominant} terms, can be explicitly given by using again the WKB approximation. This is justified since 
\be
\kappa_\rho = \sqrt{4 c^4z^2+r(z)}
\ee
obeys ${\kappa_\rho'}/{\kappa_\rho^2}\sim {1}/{z^2} \rightarrow 0$ as $z\rightarrow \infty$.
One gets
\begin{align}
    \rho(z)= w(z)e^{-\phi} &=\frac{1}{\sqrt{\kappa_\rho}} e^{-\int \kappa_\rho}\nonumber\\
    &\sim z^{-\frac{1}{2}} e^{-c^2 z^2} e^{-\frac{1}{4c^2}\int {r(z)}/{z}}.
\end{align}
The last factor captures subleading contributions in the potential; for example,\footnote{Note that in the case of the AdS metric ($w=1/z$) with quadratic dilaton $r(z)\sim a=2c^2$.} for a constant $r(z)=a$, it becomes $z^{-a/4c^2}$.
Here we ignore the exponentially growing WKB solution $\frac{1}{\sqrt{\kappa_\rho}} e^{+\int \kappa_\rho}$ due to the 
\new{problems caused by a reversed-sign dilaton discussed at the end of Sec.\ \ref{sec:setup}.}

\new{The spectrum of vector mesons can then be determined by solving}
\begin{align}
    -\chi''+(V_V(z)-E^V)\chi=0,
\end{align}
\new{which is by construction Regge-like. We define $k_V=\sqrt{E^V-V_V}$ in region II and $\kappa_V=\sqrt{V_V-E^V}$ in region IV}.

\newchange{One can also compute the decay constants of the vector mesons via $F^V_n\sim \lim_{z\rightarrow0}\rho(z) \partial_zV_n(z)$ from the mode functions $V_n(z)=\chi_n(z)/\sqrt{\rho}$. In accordance with \cite{Braga:2015jca}, $F_n^V/M_n^V$ goes to a constant as $n\rightarrow \infty$ when the potential $V_V \sim c^4 z^2$ at large $z$. (A model for $F_n^V/M_n^V$ that decays with $n\to\infty$ has been proposed in \cite{Braga:2015jca} through a modification of the holographic dictionary that would not change the divergences in the HLbL amplitude, however.)} 

The potential for the pseudoscalar and axial vector mesons
is instead given by
\begin{align}\label{eq:VAw}
    V_A=V_V+g_5^2 w(z)^2 v(z)^2,
\end{align}
\new{with $v(z)=2X_0(z)$. Let us now consider several possible cases.}

If the second term vanishes or approaches a constant as $z\rightarrow \infty$, 
one is in a similar situation as above, where even individual TFFs are not well defined. \new{In order to show this, it is necessary to be precise about the subleading terms $r(z)$.} The potential for the vector mesons reads
\begin{align}
    V_V= \frac{\kappa_\rho^2}{4}- \frac{\kappa_\rho'}{4}
\end{align}
plus terms that vanish as $z \rightarrow \infty$. Modulo such terms
\begin{align}
    V_V= c^4 z^2 +\frac{r(z)}{4}-\frac{c^2}{2}.
\end{align}

In the axial sector we \new{would then} have
 \begin{align}
     \kappa_A= \frac{\kappa_\rho}{2}-\frac{1}{z}\left(\frac{E^A}{2c^2} +\frac{1}{4}-\frac{\hat{c^2}}{2c^2} \right)+ \mathcal{O}(z^{-2}),
 \end{align}
where $\hat{c}^2$ is the limit of $g_5^2 w(z)^2v(z)^2 $
which leads to 
\begin{align}
 \label{eq:axmodfunction}
    \xi=&\frac{1}{\sqrt{\kappa_A}} e^{-\int \kappa_A}  \nonumber\\
    &\sim \sqrt{\frac{2}{\kappa_\rho}} e^{-\frac{1}{2}\int \kappa_\rho} \exp{\int \frac{1}{z} \left(\frac{E^A-\hat{c}^2}{2c^2}+ \frac{1}{4}\right) }
\end{align}
for the axial vector meson modes at $z\gg z_r$. 
\new{The function appearing in the integral yielding the TFF \eqref{eq:axtff} is}
\begin{align}
    A_n^\perp(z) =\frac{\xi_n(z)}{\sqrt{\rho(z)}}\sim z^{(E^A_n-\hat{c}^2)/2c^2}.
\end{align}
\new{By analogous reasoning one finds that the vector bulk to boundary propagator in such backgrounds still decays with a power-law at large $z$}
\begin{align}
\label{eq:vecbulktoboundgeneral}
    \J(z,Q) \sim z^{-Q^2/2c^2} ,
\end{align}
which implies that for $E^A_n \ge Q_1^2+Q_2^2+\hat{c}^2$ the TFF \eqref{eq:axtff} will not converge. 

\new{The dynamical SW model constructed in \cite{Li:2012ay,Chen:2022goa} belongs to this class of models and has $w(z)^2v(z)^2 \rightarrow0$. The earlier dynamical SW model considered in \cite{Batell:2008zm,Gherghetta:2009ac,Kapusta:2010mf} has a standard AdS warp factor $w(z)=1/z$ (in the string frame) and an asymptotically linear background $v(z)\sim z$, and thus can also not have individually well defined TFFs of the form \eqref{eq:axtff}.}

\new{Now let us consider the case when the second term in \eqref{eq:VAw} does not approach a constant as $z\rightarrow \infty$ while still being subleading compared to the first term, i.e. $\frac{w^2(z)v^2(z)}{z^2}\rightarrow 0$.
In this case 
}
\begin{align}
    \kappa_A=\kappa_V \sqrt{1+ \frac{g_5^2 w(z)^2v(z)^2}{\kappa_V^2}} \sim \kappa_V +\frac{g_5^2 w(z)^2v(z)^2}{2\kappa_V}+\ldots
\end{align}
\new{The terms that were left out need not vanish in the deep IR, but it is possible to neglect them for the following analysis. The large $z$ behavior of the axial mode function \eqref{eq:axmodfunction} gets modified by an additional factor $e^{-\int (\kappa_A-\kappa_V)}$. By assumption $\kappa_A-\kappa_V> z^{-1}$, so the new factor suppresses $A_n^\perp(z)$ at large $z$ just enough for the TFF to converge for any $E_n,Q_1,Q_2$.
Unfortunately, the exponential scaling of the TFF for large $E_n$ is still present and prevents convergence of the HLbL tensor.
In the Schrödinger frame the solution to the mode equations for region II is given by
\begin{align}
\label{eq:reg2ax}
    \xi^{\mathrm{II}} \sim \frac{1}{\sqrt{k(z)}}\cos\left( \int k(z)\right).
\end{align}
The right boundary of region II is approximately at $z=z_r \sim \frac{\sqrt{E_n}}{c^2}$. In the TFF \eqref{eq:reg2ax} gets integrated together with $\frac{1}{\sqrt{\rho}}$ (and other terms with power-law behavior at large $z$) which supplies the exponential $e^{c^2 z^2/2}$ leading to
\begin{align}
    A_n(Q_1^2,Q_2^2) \sim e^{\frac{E_n}{2c^2}}
\end{align}
for fixed\footnote{We have neither determined the power-law prefactor $(M_n^A)^\lambda$ nor the $Q_1,Q_2$ dependence, as the exponential already proves the divergence of the HLbL tensor.} $Q_1,Q_2$ by similar arguments as in section \ref{sec:TFFexpscalsubsection}.}

If the second term \new{in \eqref{eq:VAw}}
contributes at the same order or even dominates, the individual TFFs will \new{also} be well defined, but exponential scaling similar to \eqref{eq:expscalsimple} will \new{still} occur. \new{Also the axial-vector spectrum will not lie on a linear Regge trajectory (except when $w(z^2) v(z)^2\sim z^2$, but then with a different slope than vector mesons)}. The exponent \new{in the scaling law} will \new{in general also} not be proportional to $M_n^A$ anymore, rather if the second term in the \new{potential} goes like $z^\alpha$, with $\alpha\ge2$, the exponent will be proportional to $(M_n^A)^{4/\alpha}$. As $\alpha$ gets larger, the factor $e^{(M_n^A)^{4/\alpha}}$ will get smaller at fixed $M_n^A$, but as $M_n^A \rightarrow \infty$ it will still dominate any power-law factor and observables such as the light-by-light tensor will diverge.
In the case that the $w^2(z) v^2(z)$ factor behaves exponentially at large $z$, the point $z_r$ will scale as $z_r \sim \left(\log(E_n)\right)^{1/b}$, for some $b>0$. In that \new{situation}, the scaling \new{of the TFF with $E_n$} becomes a power law, and a more sophisticated analysis is needed to decide whether the infinite sums discussed in this paper converge. It seems plausible that there are ranges of $b$ for which convergence can be achieved.
Another interesting case is when $w^2(z) v^2(z)$ has a divergence at some finite $z=z_0$. In that case, $z_r$ will be equal to $z_0$ for all $n$ larger than some critical $n_c$. The exponential factor $e^{\frac{c^2}{2}z_r^2}$ is then completely harmless and the spectrum becomes hard-wall like.\footnote{If the mode functions are
nonzero for $z>z_0$, to have well defined individual TFFs one additionally needs $w(z)^2 v^2(z)\sim az^\beta$ for $z>z_0$ with $a>0,\beta \ge2$.} 
As briefly discussed at the end of the following subsection, 
the relevant sums all converge for hard-wall models. 

\new{So far, we have mostly argued in terms of the individual TFFs and their scaling with $E_n$, but one can also analyze the axial bulk-to-bulk propagator directly.

Below the diagonal $z=z'$, we have
\begin{align}
    G_A(z,z',Q)= \frac{g_5^2}{\sqrt{\rho(z) \rho(z')}} \frac{e^{\int_z^{z'} \kappa_A(\hat{z} )d\hat{z}}}{2 \sqrt{\kappa_A(z) \kappa_A(z')}},
\end{align}
where now
\begin{align}
    \kappa_A(z)=\sqrt{V_A(z)+Q^2}.
\end{align}
As before we will focus on the region close to the diagonal and parametrize $z=y+\eta$ and $z'=y- \eta$. For $y \gg\eta$ we may approximate
\begin{align}
    \int_z^{z'} \kappa_A(\hat{z} )d\hat{z} \sim -2 \eta \kappa_A(y).
\end{align}
Ignoring all factors that have a power-law scaling with $z,z'$, the axial bulk-to-bulk propagator behaves like
\begin{align}
    G_A \sim e^{c^2y^2 -2 \eta \kappa_A(y)+ \mathcal{O}(\eta^2)}.
\end{align}
At very large $y$, we may safely neglect the $\mathcal{O}(\eta^2)$ terms if $\eta$ is smaller than some fixed constant $\delta$.
To compute the coefficients of the HLbL tensor $\hat{\Pi_i}$ one then needs to multiply by several vector bulk-to-boundary propagators $\J$ (whose large $z$ behavior was found to be power law in \eqref{eq:vecbulktoboundgeneral}) and integrate over $z,z'$. 
The $d \eta$ integral over a small strip of width $\delta$ can then be simplified to
\begin{align}
    \int dy\, e^{c^2 y^2} p(y)\frac{(e^{- \delta  \kappa_A(y)}-1)}{-2 \kappa_A(y)} \rightarrow   \int dy\, e^{c^2 y^2} \tilde{p}(y) 
\end{align}
where $p(y),\tilde{p}(y)$ collect terms scaling with a power law in $y$. This clearly diverges \emph{irrespective} of $\kappa_A(z)$ and by extension $V_A(z)$ (provided they are not singular at some finite $z$)! The five-dimensional computation thus confirms the conclusions of the earlier analysis done in terms of TFFs. The only assumptions that entered, besides the field content (and existence of the terms \eqref{eq:actionassump}), were the Regge behavior of the vector meson spectrum, regularity of the backgrounds at all finite $z$, and the presence of a Chern-Simons term of the form \eqref{SCS}.
}

\subsection{Scalar-extended CS term}

Seemingly the only way to deal with the exponential scaling \eqref{eq:TFFscaling} and the related convergence issues without adding new fields or introducing backgrounds with singularities \new{at finite $z$}, \new{while retaining Regge behavior of vector mesons,} is by \new{modifying the Chern-Simons term.
In fact, a natural modification motivated by string theory involves}
grouping together the gauge fields $A_L,A_R$ and the bifundamental scalar $T \sim X$ into a superconnection \cite{Quillen:1985vya}, leading to a scalar-extended Chern-Simons term.

Such a Chern-Simons term is generated in string theory when a brane and an anti-brane come close to each other. The string stretching between the two branes gives rise to the bifundamental scalar and is tachyonic. (There are no problems with unitarity if the background is AdS and the square of the mass is above the Breitenlohner-Freedman (BF) bound \cite{Breitenlohner:1982bm}.)

In the (scalar extended) Chern-Simons term of this superconnection there now appear factors of $e^{-2 \pi \alpha' T_0^2}$ that can regulate the divergences of the previous sections. 

The \new{pseudoscalar} transition form factor has been worked out in \cite{Leutgeb:2024rfs} and reads (in the $a=3$ sector)
\begin{align}
\label{eq:finpstff}
    F(Q_1^2,Q_2^2)&=\frac{N_c}{2 \pi^2} \int dz \,\text{tr}\, \Q^2 t^3 \bigg\{e^{-2 \pi \alpha' T_0^2}\varphi'\J_1 \J_2 \nonumber \\ &+(e^{-2 \pi \alpha' T_0^2})'(\varphi-\eta)\J_1\J_2\bigg\},
\end{align}
where $\alpha'=\frac{1}{6}$, $\J_i=\J(z,Q_i)$ and the tachyon $T_0$ is proportional to $X_0$.
\new{The axial TFF obtained from the scalar extended Chern-Simons term reads
\begin{align}
    \label{eq:axtffscalex}
    A_n(Q_1^2,Q_2^2)=\frac{2}{Q_1^2}\int\;  dz \;\text{tr}(t^3 \Q^2 e^{-2 \pi \alpha' T_0^2} )  A^{\perp}_n(z) \J_1' \J_2.
\end{align}
}
If the tachyon $T_0$ is identified with $X_0$ using the Dirac-Born-Infeld (DBI) action of the brane-antibrane system, one has \cite{Leutgeb:2024rfs} 
\begin{align}
\label{eq:tachidentific}
    T_0^2=\frac{g_5^2 \pi}2 X_0^2
    =\frac{g_5^2 \pi}8 v^2.
\end{align}

In the standard SW model of the previous section {$v^2 \sim \sigma^2 z^6$}, so the factor $e^{-2 \pi \alpha' T_0^2}$ would introduce exponential damping that suppresses the $e^{\frac{c^2z^2}{2}}$ factor from the dilaton at large $z$. The exponential scaling then does not occur, and a finite HLbL tensor is obtained.

If, however, the background is given by a regular $X_0$ that goes to a constant (\new{as in \cite{Li:2012ay,Chen:2022goa}, where it goes to zero}), then this scalar extension does not help \new{(and, as shown before, not even individual TFFs are well defined if in addition $w(z)^2 v(z)^2 \rightarrow const$). Hence,} an unbounded $X_0(z)$ background seems to be necessary. 

\new{Interestingly enough, an unbounded $X_0(z)$ means that chiral symmetry is not restored in highly excited hadrons, and arguments in favor
of this scenario have been put forward in \cite{Shifman:2007xn}. 
}

\new{In the model of \cite{Batell:2008zm,Gherghetta:2009ac,Kapusta:2010mf} the background $v(z)\sim z$, which together with the scalar extended Chern-Simons term makes the integral yielding individual TFFs \emph{finite}. }

\new{In order to avoid exponential scaling of the TFF with $n$ that prevents
the existence of the complete HLbL amplitude, the factor $e^{-2 \pi \alpha' T_0^2}$ needs to be able to suppress the exponential factor $e^{c^2z^2/2}$ coming form $\frac{1}{\sqrt{\rho}}$. In other words, we need
\begin{align}
    \frac{2 \pi \alpha' T_0^2}{z^2}>\frac{c^2}{2}
\end{align}
as $z\rightarrow \infty$. The integral yielding the TFF is then damped at large $z$, which gets rid of the exponential scaling with $E_n$. Particularly appealing is the case when at the same time ${w^2 v^2}/{z^2}\rightarrow const$ (or $0$), since then as discussed above also the spectrum in the axial sector
is Regge-like.

In the special case $\frac{v^2}{z^2}\rightarrow const$ considered in \cite{Batell:2008zm,Gherghetta:2009ac,Kapusta:2010mf}, 
whether the HLbL amplitude is finite or not depends on the identification \eqref{eq:tachidentific} and the coefficient $A$ in $v(z) \sim A\, z$.
}

\new{Coming back to the conventional SW model considered in the previous section
(where only the vector meson spectrum has Regge behavior),
we see that the convergence issue in the HLbL amplitude can be fixed by
a scalar-extended Chern-Simons term. However, the
phenomenology obtained with \eqref{eq:tachidentific}
is even more unsatisfactory for the neutral $\pi^0$
than in \cite{Colangelo:2023een}: the TFF at large but not asymptotically large  $Q^2$ is more strongly overestimated before the
Brodsky-Lepage limit is eventually approached from above.}
The contribution $a_\mu^{\pi^0,\mathrm{SW}}\approx 75 \times 10^{-11}$
becomes $89 \times 10^{-11}$ 
with the above scalar-extended Chern-Simons term\footnote{The parameter $\alpha'$ is set to $\frac{1}{6}$ in units of the AdS radius to have \new{a tachyon mass corresponding to} the correct scaling dimension of the quark condensate operator in the boundary theory. \new{It is thus in tension with
the supergravity limit, which assumes $\alpha'\to0$}.},
even further above the current dispersive estimates of $62.6 \times 10^{-11}$ \cite{Hoferichter:2018kwz}.

Hard-wall models do not suffer from the above exponential scaling of TFFs since the modes are restricted to a finite region $z\in[0,z_0]$.\footnote{This is also the case for the Witten-Sakai-Sugimoto model when brought into Schrödinger form. There, however, the geometry is not asymptotically AdS so that one cannot match to the short-distance constraints of QCD.} This leads to the scaling behaviors $F(0,0) \sim E_n^{-1}, f_n\sim E_n^{-1/4}$ with $E_n \sim n^2$, which leads to a convergent sum rule \eqref{eq:sumrule}, and a convergent light-by-light tensor for the pseudoscalars (the axials are analogous).
\new{Moreover, the experimental data for the pion TFF (and also for the $a_1$) turn out to be
reproduced surprisingly well \cite{Leutgeb:2022lqw,Ludtke:2024ase,Aliberti:2025beg}
with a standard Chern-Simons term.\footnote{In the HW model, a scalar-extended Chern-Simons
term with $\alpha'=\frac16$ helps to improve some features in the $f_1$-$f_1'$
sector, but also leads to an unsatisfactory pion TFF \cite{Leutgeb:2024rfs}.}
}

\subsection{V-QCD}

The holographic V-QCD framework \cite{Casero:2007ae,Gursoy:2007cb,Gursoy:2007er,Iatrakis:2010zf,Iatrakis:2010jb,Jarvinen:2011qe}
aims to describe QCD in the Veneziano limit and 
\new{is motivated by general properties of} systems of $N_f$ branes and antibranes in backgrounds sourced by $N_c$ color branes in string theory. There is considerable freedom to add potentials in this framework, which can then be fitted to experimental and lattice data of QCD.
In V-QCD, the backreactions are taken into account, and the coupled non-linear equations of motion are typically solved numerically. However, for large and small $z$, analytic expressions for the various background solutions are available. They depend, of course, on the specific potentials used in the action. \new{V-QCD models do not fall into the rather general class of models considered in the previous subsection because the action analogous to \eqref{eq:actionassump} is slightly more complicated, and therefore, these models need to be analyzed separately.} 
 
With regard to transition form factors, the situation in V-QCD will be very similar to the simple example from the previous section with the scalar extended Chern-Simons term and the unbounded $X_0$. The main difference is that in V-QCD all the background fields are bona fide solutions to the equations of motion.

Here we will consider the concrete choice of potentials in \cite{Jarvinen:2022gcc} and focus on the pseudoscalar sector.
The action (in their \new{notation and} mostly + convention) reads\footnote{The coordinate $z$ here is called $r$ in \cite{Jarvinen:2022gcc}; the field $\pi(z,y)$ in \cite{Jarvinen:2022gcc} corresponds to $\frac{1}{2}\eta(z,x)$.\new{ The function $w$ in \eqref{eq:vqcdaction} is \emph{not} related to the warp factor, which in this equation is given by $e^A$.}}
\begin{align}
\label{eq:vqcdaction}
   S= \int d^4x dz \; V_f e^{A}G^{-1} \bigg\{  \tau_0^2\kappa e^{2A}(\partial_z \eta +2A_z)^2\nonumber\\+ G^2\tau_0^2\kappa e^{2A}(\partial_{\mu} \eta +2A_{\mu})^2  +w^2(\partial_z A_{\mu}-\partial_{\mu} A_z)^2 \bigg\} 
    \end{align}
with 
\begin{align}
    G&=\sqrt{1+\kappa e^{-2A}(\partial_z\tau_0)^2 } \nonumber \\
    A_{\mu}&=-\partial_{\mu} \varphi.
\end{align}
{The potential $V_f$ depends on the tachyon in the following way}
\begin{align}
    V_f=V_f^0(\lambda) e^{-a_{IR} \,\tau_0^2}.
\end{align}
\new{The quadratic action is augmented by a scalar-extended Chern-Simons action generalizing further the one given above \cite{Jarvinen:2022mys}.} {A parameter $b$ is added which introduces exponential damping for the Chern-Simons term of the form}
\begin{align}
    S_{CS} \sim \int e^{-b \,\tau_0^2}\times \left(\text{5-forms}\right).
\end{align}
In order to transform the equations of motion 
into a Schrödinger-type form one needs in addition to redefinitions of $\pi, \varphi$, a variable transformation $x(z)$ with
\begin{align}
    \frac{\partial x}{\partial z}=G.
\end{align}

The precise definitions and asymptotics for the potentials can be found in Appendix B of \cite{Jarvinen:2022gcc}.
Let us note that the background of the bifundamental scalar, \new{the tachyon}, which is denoted here as $\tau_0$, is unbounded and goes to $\infty$ as $z\rightarrow \infty$.
It can be checked that the point $x_r$ where $E_n-V=0$ scales like $E_n^{1/2}$; hence, Regge behavior $E_n \sim n$ is realized.
The transition form factors in this model are well defined for any individual mode since V-QCD uses a scalar-extended Chern-Simons term \new{with exponential suppression at
large $z$ due to the unbounded $\tau_0$}.

The most important point that we want to check is whether the exponential scaling behavior of TFFs occurs in this model.
The TFF in V-QCD is given by an integral over $$\exp \left\{-(2/3)^{\frac{1}{2}}\frac{1}{\kappa_\mathrm{IR}} (b-\frac{a_\mathrm{IR}}{2})x^2 \right\}$$ times an oscillating function of period $\sqrt{E_n}$ with an average height that has a power-law behavior in $E_n$.
The parameter $a_\mathrm{IR}$ parametrizes the tachyon potential $V_f$ in the DBI action, while $b$ appears in their scalar-extended Chern-Simons term.
The TFF will have exponential scaling behavior similar to \eqref{eq:TFFscaling} if $b-\frac{a_\mathrm{IR}}{2}<0$.

{In V-QCD one typically already chooses $b>a$ to obtain regular background solutions in the case of a nonzero $\theta$ angle \cite{Ishii:2019gta,Arean:2016hcs}.
The issues of convergence discussed in this paper are thus avoided.}

As mentioned before, regarding the exponential scaling of the transition form factors, the V-QCD model is very similar to the simple soft-wall model with scalar-extended Chern-Simons term and unbounded $X_0$ discussed earlier.
V-QCD in addition has backgrounds consistent with the equations of motion and Regge behavior of the masses also in the pseudoscalar and axial-vector sector.
The exponential suppression in the scalar-extended Chern-Simons term also saves the HLbL tensor in V-QCD.

\section{Conclusion and discussion}

In this paper we have uncovered divergences in infinite mode sums 
for observables related to the Chern-Simons term in various soft-wall models. We have also shown that in certain 
cases such as the VVA correlator the corresponding $5$-dimensional calculation gives a perfectly well-defined result, and we have pinpointed the divergence in the mode sum to a non-commutativity of $\int_0^{\infty} dz$ and $ \sum_{n=0}^{\infty}$.

The hadronic light-by-light tensor is, however, divergent 
even when directly calculated from the 5-dimensional framework
with bulk-to-bulk propagators.
We therefore conclude that the conventional SW model as employed in
\cite{Colangelo:2023een,Colangelo:2024xfh}
is unsuitable for an evaluation of the hadronic light-by-light contribution to the muon $g-2$.
Unlike the HW models studied previously,
the SW model fails to provide a hadronic model where
the MV-SDC can be satisfied.\footnote{Ref.\ \cite{Colangelo:2024xfh}
reports that the MV-SDC was satisfied in the SW model. This will indeed
be the case when the SW model is evaluated with a fixed cutoff on $z$.
However, the numerical results for the HLbL amplitude as well as the
$a_\mu$ contribution of the tower of resonances in the axial sector
will depend on the cutoff, yielding infinity when the cutoff is removed.}
Even when used only as a model for the lowest excitations,
the SW model does not appear to be in good shape.
As noted already in \cite{Colangelo:2023een},
the SW model is in poor agreement for the ground-state
pion contribution. Experimental and dispersive results for
the $\pi^0$ TFF are not well reproduced, and $a_\mu^{\pi^0}$
is clearly overestimated when compared to dispersive and
lattice results. On the other hand, the HW model introduced in \cite{Erlich:2005qh} as well as variants thereof
as studied in \cite{Leutgeb:2021mpu}
agree remarkably well with phenomenological results.
Similarly, the $a_\mu$ contribution from the $a_1$ axial-vector meson is larger in the SW model than in the HW model,
whose prediction for 
the singly virtual TFF predicted by the latter
for the $a_1$ has been found to agree surprisingly well
with the recent dispersive results of \cite{Leutgeb:2022lqw,Ludtke:2024ase,Aliberti:2025beg}.

Regarding excited states, it would certainly be interesting
to consider holographic models that have more realistic
Regge trajectories than the HW model while avoiding
the pathologies of the standard SW model and some of its generalizations.
As we have seen, those can be cured by a scalar-extended CS
term in the standard SW model (though not in the model of
\cite{Li:2012ay,Chen:2022goa}).
However, this is not well motivated phenomenologically, since
the discrepancy with data-driven results for $a_\mu^{\pi^0}$
is even more severe than in the standard SW model of \cite{Colangelo:2023een}.
A more promising class of models 
(albeit with a rather large parameter space)
would be V-QCD, where
in addition to a scalar-extended CS term the
background geometry is modified as well, ensuring
also correct asymptotic Regge behavior in the axial sector.

\section*{Acknowledgments}
We would like to thank Luigi Cappiello, 
Matti J\"arvinen, Elias Kiritsis, 
and Harald Skarke for helpful discussions.
This work has been supported by the Austrian Science Fund FWF, Grant-DOI
10.55776/PAT7221623.

\appendix
\section{Details on the WKB approximation}
\label{app:WKBDetails}

In this appendix we \new{recapitulate} the WKB approximation and use it to find the pseudoscalar and axial vector mode functions for high masses in the minimal soft-wall model described in the introduction.

For equations of the form
\begin{align}
    y''+k^2(z)y=0
\end{align}
one can make the transformation $y=e^{i \Phi}$ leading to
\begin{align}
    \Phi'=\sqrt{k^2+i \Phi''}.
\end{align}
and attempt to solve it at large $k^2$. For simplicity we assume $k>0$ here.
At large $k^2$ the zeroth order solution is $\Phi_0= \int k$.
Upon reinserting it into the above equation, one sees that the zeroth order solution is a good \new{first} approximation if $|k'| \ll k^2$. 

Equations of the form $\Phi'=f(z,\Phi)$ may be reformulated as fixed points of 
\begin{align}
    \Phi_n(z)= \Phi_{n-1}(z^*)+\int_{z^*}^z dz' f(z,\Phi_{n-1}).
\end{align}
With suitable starting values $\Phi_0(z)$, the limit $\Phi_n(z)$ will converge to a solution of the original differential equation. For our purposes we may use an approximation to $\Phi_1$, 

which follows from $\int dz' \sqrt{k^2+i\Phi_0''}\approx
\int dz' k + \frac{i}{2}\int dz' \Phi_0''/k=\int dz' (k+\frac{i}{2}\frac{k'}{k})$ and
which in terms of $y$ reads
\begin{align}
    y=\frac{1}{\sqrt{k(z)}}(A e^{i\int k}+B e^{-i\int k}).
\end{align}
when $k(z)^2$ is positive. If $k^2=- \kappa^2$ is negative, the appropriate solution is
\begin{align}
     y=\frac{1}{\sqrt{\kappa(z)}}(A e^{\int \kappa}+B e^{-\int \kappa}).
\end{align}
Around the regions where $k(z)^2=0$ the true solution will not be well described by the WKB expression as it diverges there.

\subsection{Pseudoscalar modes}
In the following part, we spell out some of the details omitted in the main part of the paper concerning the WKB approximation for the pseudoscalar sector. Axial vectors will be investigated in the next subsection.
The Schrödinger equation for the pseudoscalar modes with $k^2(z)=E-V_P(z)$ is amenable to a WKB analysis for generic values of $z$. In this appendix, we will spell out in which regions the WKB approximation fails and which approximation to use instead. We will also find the points $z_i$ at which the solutions to the right and left are both valid and can therefore be matched at large $E_n$.

Since the potential is bounded from below and the solution needs to be normalizable, we expect that solutions exist only for discrete values of $E=E_n$.
The WKB approximation breaks down when $\frac{k'}{k^2} \ll 1$ is violated. For any given fixed $z$, one may observe that $\frac{k'}{k^2} \rightarrow 0$ as $E_n\rightarrow \infty$.
For any large but fixed $E$, there are, however, always regions that are not described well by the WKB approximation.
These are the regions around the two zeros $z_l,z_r$ of $k^2(z)$ which at large $E_n$ are given approximately by 
\begin{align}
    z_l&=\sqrt{\frac{15}{4 E_n}},\nonumber\\
    z_r&=\bigg(\frac{E_n}{g_5^2 \sigma^2}\bigg)^{\frac{1}{4}},\nonumber
    \label{eq:zrApp}
\end{align}
as well as the very small-$z$ region where $\kappa^2 \approx \frac{15}{4z^2}$ and  therefore $\frac{\kappa'}{\kappa^2}\sim \frac{1}{z^4}$.
In these regions, the differential equation needs to be solved with different methods.
If we pick $z_1(E_n)$ in such a way that it approaches $0$ in the limit, then we may always approximate $\kappa^2=E_n-V_P\sim E_n-\frac{15}{4z^2}$ in the whole range $z \in [0,z_1]$ at large $E_n$. With this form of $\kappa^2$, the analytic form of the solution that is normalizable is \eqref{eq:BesselJtext}. Clearly, it is also a good approximation at $z=z_1$. We however do not want to let $z_1(E_n)$ go to zero too quickly. The point $z_l$ at which the WKB solution diverges goes to zero like $E_n^{-\frac{1}{2}}$ and $z_1$ should always be larger than $z_l$ in order to have a chance of the WKB approximation being valid at $z_1$.
Computing $\frac{\kappa'}{\kappa^2}$ at $z=z_1$ with the ansatz $z_1 = z_l \left( \frac{E_n}{c^2}\right)^{a_1}$ and demanding it to go to zero as $E_n \rightarrow \infty$ reveals
\begin{align}
   0< a_1.
\end{align}
Since $z_1 \rightarrow 0$, obviously
\begin{align}
    a_1<\frac{1}{2}
\end{align}
has to hold. This explains the form of $z_1$ mentioned in the main text.

The second point $z_r$, where $k^2=0$ requires a different approach. Large $E_n$ implies large $z_r$, and hence, the potential is approximated by $V_P\sim g_5^2 \sigma^2 z^4$ around this point. Around $z=z_r$ one may use the series expansion 
\begin{align}
    k^2=E-V_P \sim -(z-z_r)V'|_{z=z_r}.
\end{align}
Neglecting the higher order terms is justified when $|(z-z_r )\frac{V''}{V'}|\ll 1$ that is when $|\frac{z-z_r}{z_r}|\ll 1$.
This prompts us to make the ansatz
\begin{align}
    z_2=z_r\left(1-\left(\frac{E_n}{c^2}\right)^{-a_2}\right)\\
     z_3=z_r\left(1+\left(\frac{E_n}{c^2}\right)^{-a_2}\right)
\end{align}
where $0<a_2$, which mark the boundaries of the region where we can linearize $k^2$. In principle it would have been possible to introduce two different exponents or to introduce numerical constants in front of the $\frac{E_n}{c^2}$ term. This is, however, completely unnecessary.
Within the region $[z_2,z_3]$, the mode equation with linearized potential has an exact solution
\begin{align}
\label{eq:Airy III}
    y(z)=& C_3 \Ai\left(L^{\frac{1}{3}}(z-z_r)\right),
\end{align}
where $L=V'|_{z=z_r}$.
As before we would not like to make $a_2$ too big because then $z_2,z_3$ would go too quickly toward $z_r$, causing the WKB approximation to not be valid anymore at $z=z_2,z_3$ (at $z_2$ it is the oscillating WKB solution of region II, at $z_3$ it is the exponentially decaying WKB solution of region IV). A short computation of $\frac{k'}{k^2}$ at $z=z_2,z_3$ reveals that $a_2<\frac{1}{2}$ so that it goes to zero as $E_n \rightarrow\infty$.

Matching the different solutions at the points $z_i$, one can fix all but an overall constant (that we choose to be $C_1$). Since we have already thrown away non-normalizable solutions, the allowed $E_n$ will form a discrete set. 
The overall constant $C_1$ will then be determined by demanding the solution to obey the norm condition \eqref{eq:normconps}.

The relations that arise from the matching of solutions, which is done at $E_n = \infty$, are
\begin{align}
\label{eq:coefstart}
        C_2&=-\frac{C_1}{\sqrt{2 \pi}}\exp (i\frac{\sqrt{15}}{4}\pi -i\frac{\pi}{4}) \\
        \label{eq:coefsecond}  
    D_2&=-\frac{C_1}{\sqrt{2 \pi}}\exp (-i\frac{\sqrt{15}}{4}\pi +i\frac{\pi}{4})
    \end{align}
        \begin{align}    
    C_2&\exp [i \int_{z_l}^{z_r} k(z') dz']= C_3 \exp (i \frac{\pi}{4}) \frac{1}{2 \sqrt{\pi}} L^{1/6}\\
      D_2&\exp [-i \int_{z_l}^{z_r} k(z') dz']= C_3 \exp (-i \frac{\pi}{4}) \frac{1}{2 \sqrt{\pi}} L^{1/6}
      \end{align}
      \begin{align}
       C_4&= C_3  \frac{1}{2 \sqrt{\pi}} L^{1/6}  \label{eq:coefend}
        \end{align}
      \begin{align}
       & \int_{z_l}^{z_r}k(z)dz- \frac{\pi}{2}+ \frac{\sqrt{15}}{4}\pi=n \pi.
\end{align}
The last equation gives a constraint on the allowed $E_n=m_n^2$. The integral can be computed approximately for large $E_n$ as
\begin{align}
    \int_{z_l}^{z_r}k(z)= \frac{E_n^{3/4}}{\sqrt{g_5 \sigma}}\int_0^1 \sqrt{1-u^4}du = \frac{2}{3}\frac{E_n^{3/4}}{\sqrt{g_5 \sigma}}K(-1),
\end{align}
where $K(x)$ is the complete elliptic integral of the first kind. Therefore $m_n^2 \sim n^{4/3}$.  The prefactor $E_n^{\frac{3}{4}}$ arises from the product of a factor of $z_r$ coming from transforming the variables $z(u)= z_r\, u$ and a factor $\sqrt{E_n}$ from $\sqrt{E_n-V_P}$.
Only if $z_r \sim E_n^{\frac{1}{2}}$, or $V_P \sim z^2$ at large $z$ can Regge behavior for pseudoscalars arise \new{at large $n$}. 

The integral in the calculation of the norm can be split up into integrals over the four regions, where the contribution from region \textrm{II} is dominant in the large $E_n$ limit.
From this region we get (after a variable redefinition)
\begin{align}
      1&=C_1^2 E_n^{3/4}\frac{2 }{\pi g_5^2} \frac{1}{\sqrt{g_5 \sigma}} \nonumber \\ &\times\int_0^1 \frac{1}{\sqrt{1-u^4}} \cos^2 \bigg(\frac{E_n^{3/4}}{\sqrt{g_5 \sigma}}\int_0^u \sqrt{1-u'^4}du'\bigg)
\end{align}
Since the square of the cosine is oscillating very rapidly we may approximate it by $\frac{1}{2}$, leading to
\begin{align}
\label{eq:normpseudo}
    1=  \frac{K(-1)}{\pi g_5^2} \frac{1}{\sqrt{g_5 \sigma}}  C_1^2 E_n^{3/4}.
\end{align}
The solution has thus been fully determined.

As said before, we have taken the points $z_i$ to scale with $E_n$ in such a way that in the large $E_n$ limit, the solutions to the left and to the right of them are valid and converge toward each other at $z_i$ as $E_n \rightarrow \infty$. At any finite $E_n$, there is, however, always a finite difference between these two solutions at $z_i$. This can be seen in figure \ref{fig:comparison} at the points indicated in the caption. As we will later be interested in quantities where $y$ is integrated, these finite jumps (that disappear for $E_n \rightarrow \infty$) will play no role.

In principle, one may find finite regions around the $z_i$ and glue the two solutions in a more smooth manner, which may improve numerical results at low mode numbers. For our purposes this is unnecessary though.

\subsection{Axial vector modes and nonzero quark mass}
Since the potential for the axial vector mesons is qualitatively very similar to the pseudoscalar potential, the above analysis applies almost immediately to this case as well.
The point $z_l$ is now approximately at
\begin{align}
 z_L=   \sqrt{\frac{3}{4 E_n}}
\end{align}
and in region \textrm{I} the solution changes to
\begin{align}
  \xi= C^A_1  \sqrt{z} J_1(\sqrt{E_n}z)
\end{align}

In the relations \eqref{eq:coefstart}-\eqref{eq:coefend} obtained from matching  one has to change \eqref{eq:coefstart} and \eqref{eq:coefsecond} to
\begin{align}
    C_2^A&=-\frac{C_1^A}{\sqrt{2 \pi}} \exp (i \frac{\sqrt
    3}{4}\pi +i \frac{\pi}{4}),\\
     D_2^A&=-\frac{C_1^A}{\sqrt{2 \pi}} \exp (-i \frac{\sqrt
    3}{4}\pi +-i \frac{\pi}{4}).
\end{align}
The relation determining the energies then changes to
\begin{align}
    \int_{z_l}^{z_r} k(z) dz +\frac{\sqrt{3}}{4} \pi=n \pi,
\end{align}
while the rest of the relations remains the same. The quantity $k(z)=\sqrt{E_n-V_A(z)}$ is defined here using the \emph{axial} Schrödinger potential, and similarly for $\kappa(z)$. 
Asymptotically, the allowed $E_n$ are therefore the same as in the pseudoscalar case and the computation of the norm proceeds analogously, which means
\begin{align}
      1&=(C^A_1)^2 E_n^{-1/4}\frac{2 }{\pi g_5^2} \frac{1}{\sqrt{g_5 \sigma}} \nonumber \\ &\times \int_0^1 \frac{1}{\sqrt{1-u^4}} \cos^2 \bigg(\frac{E_n^{3/4}}{\sqrt{g_5 \sigma}}\int_0^u \sqrt{1-u'^4}du'\bigg) \\
    \label{eq:axnorm}  &=  \frac{K(-1)}{\pi g_5^2} \frac{1}{\sqrt{g_5 \sigma}}  (C_1^A)^2 E_n^{-1/4}.
\end{align}

Axial-vector decay constants are given by
\begin{align}
    F_A=-\frac{1}{g_5^2} \frac{\partial_z A^{\perp}}{z}\bigg|_{z=0}= -\frac{C_1^A}{g_5^2}\sqrt{E_n}.
\end{align}
To compute pseudoscalar decay constants, we need to leave the chiral limit, which complicates the analysis.\footnote{The ground state $\pi^0$ has a nonvanishing $f_\pi$ also in the chiral limit, but for all excited pions $f_{_n}\sim m_q$.}
For now let us assume $m_q$ is much smaller than any dimensionful constant in the theory. The small-$z$ behavior of the potential $V(z)$ then changes completely. It will go to $-\infty$ instead of $+ \infty$, and for large enough $E_n$, there will be no more $z_l$ at which $E_n-V=0$ (the point $z_r$ remains present and is given to leading order by the same formula as in the chiral case). The small-$z$ expansion of the potential reads
\begin{align}
    V_P= -\frac{1}{4 z^2}-\frac{4\sigma}{m_q}+ O\bigg(\frac{z}{m_q}, \frac{1}{m_q^2}\bigg).
\end{align}
With this potential the condition for the validity of the WKB approximation $\frac{k'}{k^2}\ll1$ is obeyed down to $z_1= E_n^{-\frac{1}{2}} (E_n c^{-2})^b$ with $0<b<\frac{1}{2}$ which in the large $E_n$ limit is much smaller than $\frac{1}{m_q}$. The appropriate solution for $[0,z_1]$ can be found analytically and reads
\begin{align}
  y=  C_1 \sqrt{z} J_0(\sqrt{E_n}z)+ D_1\sqrt{z}Y_0(\sqrt{E_n}z)
\end{align}
The second term would give an infinite contribution to $f_{\pi}$ so we set $D_1=0$.
The pion decay constant reads
\begin{align}
    f_{\pi}= -\frac{m_q}{g_5}\frac{1}{\sqrt{z}}y\Big|_{z=0}= -\frac{m_q }{g_5}C_1
\end{align}
After redoing the matching to the WKB solution in region \textrm{II}, one can see that to leading order in $m_q$ and $E_n$ the norm does not change and $C_1$ is still \new{determined} by \eqref{eq:normpseudo}.

\section{Transition form factors of HW models in WKB approximation}
\label{app:HWTFF}
\begin{figure}
    \centering
    \includegraphics[width=1\linewidth]{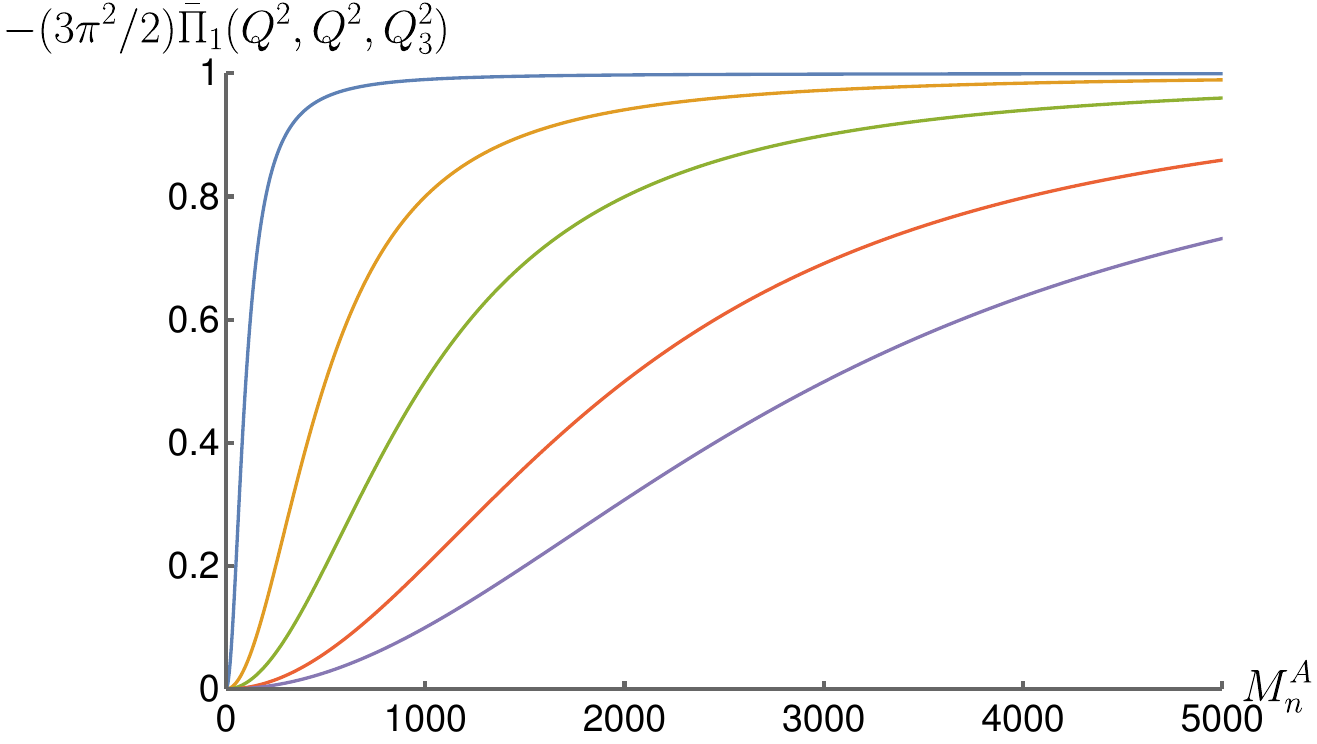}
    \caption{Summing the contributions to the MV constraint from modes with masses less than $M^A_n$ {(in units of GeV)} for various $Q_3$ and $Q=30 000$ GeV shows that it is mostly the modes with $M_n^A\sim Q_3$ that contribute. The curves for $Q_3=100,\,500,\,1000,\,2000, \,3000$ GeV in blue, orange, green, red and purple indicate that the most important contributions come from the regime $M^A_n\sim Q_3$ where the asymptotic expansions of \cite{Hoferichter:2020lap} are not relevant. }
    \label{fig:MVc}
\end{figure}

One may also use the WKB approximation to give explicit expressions for the TFFs at high energies $Q$ in hard wall models. These are particularly interesting in the axial sector since they are responsible for satisfying the Melnikov-Vainshtein constraint.
\new{For $Q^2\gg m_\rho^2, m_q^2, \sigma^{2/3}$ but arbitrary $E_n$, one can derive the following} analytic expressions in the singly virtual and doubly virtual cases:
\begin{align}
    A_n(Q^2,0)=-\frac{2C_1^A}{Q^2} \text{tr}(t^3 \Q^2  )\frac{2 \sqrt{E_n}Q^2}{(E_n+Q^2)^{2}},\\
     A_n(Q^2,Q^2)=-\frac{2C_1^A}{Q^2}\text{tr}(t^3 \Q^2  )\frac{2}{E_n(E_n+4Q^2)^{\frac{5}{2}}} \nonumber \\
    \times \bigg(Q^2(E_n-2Q^2)\sqrt{E_n(E_n+4Q^2)}\nonumber\\+8Q^4(E_n+Q^2)\asinh \big(\frac{\sqrt{E_n}}{2 Q}\big)\bigg).
\end{align}
where $C^A_1=-g_5^2 \frac{F_A}{\sqrt{E_n}}$ can be computed using \eqref{eq:axnorm}.
These should hold for all hard-wall models, irrespective of boundary conditions.
The expansions for $Q\gg \sqrt{E_n}$ of the above expressions are fully consistent with \cite{Hoferichter:2020lap}. In terms of the mode number $n$, we find
\begin{align}
      F_A^n&=\sqrt{\frac{\pi}{z_0}}\frac{E_n^{\frac{3}{4}}}{g_5}\\
    E_n&=\frac{\pi^2}{z_0^2}n^2.
\end{align}
By explicit numerical resummation, one may check that the MV constraint is satisfied and that for a given large $Q_3, Q$ with $Q_3\ll Q$, it is the modes $\sqrt{E_n}\sim Q_3$ that give the bulk of the contribution. As indicated in figure \ref{fig:MVc}, the modes with $\sqrt{E_n}\gg Q_3$ contribute only negligibly.

In the symmetric limit, it is known that hard-wall holographic models obtain the correct scaling, but the axial vectors reproduce only about $81$\% of the coefficient, which may also be verified numerically using the formulas above. Recently it was shown in \cite{Mager:2025pvz,Cappiello:2025fyf}, that the remaining $19\%$ of the symmetric SDC can be matched by tensor mesons. 

\section{HLbL amplitude in the Witten-Sakai-Sugimoto model}

In this appendix we investigate HLbL scattering and potential convergence issues in the Witten-Sakai-Sugimoto (WSS) model, which in
contrast to the bottom-up models is based on a string-theoretic construction. While this model cannot be matched to the
short-distance constraints of QCD and correspondingly is
less interesting phenomenologically, regarding finiteness, the WSS model behaves essentially like a hard-wall model and no divergences are found. 

As the tower of excited pions is absent in the WSS model, the contribution to the HLbL tensor in the pseudoscalar sector is given solely by the ground-state Goldstone bosons and is thus trivially finite. The model does however contain an infinite number of excited axial vector mesons, which we now analyze.

Both vector and axial vector meson modes are described by a $U(N_f)$ gauge field $A$ on the flavor branes whose holographic direction\footnote{We do not consider fluctuations on the $S^4$ part of the branes.} is described by the coordinate $Z \in(-\infty,\infty)$. They are encoded in the even and odd parts of $A_\mu$, respectively. We therefore work on the half interval $Z\in [0, \infty]$.
The vector and axial vector equations of motion are identical
\begin{align}
\label{eq:EOMSS}
  \partial_Z(K \partial_Z\mathcal{J})-K^{-1/3} \left(\frac{Q}{M_\mathrm{KK}}\right)^2 \mathcal{J}=0, \\
    \partial_Z(K \partial_Z\psi_A)+K^{-1/3} \left(\frac{M^A}{M_\mathrm{KK}}\right)^2 \psi_A=0,
\end{align}
where $K=1+Z^2$, but differ in the IR boundary conditions
\begin{align}
    \J'(0,Q)=0, \nonumber \\
    \psi_A(0)=0.
\end{align}
To apply the WKB machinery we transform the equations of motion into the form of a Schrödinger equation using the variable transform
\begin{align}
     \frac{\partial x}{\partial Z}=K^{-2/3},
\end{align}
which is solved by
\begin{align}
\label{eq:coordtraf1SS}
    x(Z)=Z\,{}_2F_1(\frac12,\frac23;\frac32,-Z^2).
\end{align}
This transformation maps the non-compact $Z$ interval to a compact interval $x \in[0,x_f]$ with $x_f=x(Z=\infty)=\frac12 B(\frac12,\frac16)$ and $B(a,b)$ the Euler beta function. With respect to these coordinates the IR boundary is at $x=0$ and the UV boundary at $x=x_f$.
We then define $\xi$ via
\begin{align}
    \psi_A= \frac{1}{\sqrt{\rho}} \xi, \qquad \rho=K^{1/3},
\end{align}
and similarly for the vector sector, and in addition transform variables one more time to
\begin{align}
\label{eq:coordtraf2SS}
    z=z_0-x,
\end{align}
where $z_0=x_f$ to arrive at 
 \begin{align}
 \label{eq:SSschroed}
     -\partial_z^2\xi+ V_\mathrm{WSS}\,  \xi=E \,\xi,
 \end{align}
with $E= (M^A/M_\mathrm{KK})^2$ or $E=-(Q/M_\mathrm{KK})^2$.
Just like in HW models, the asymptotic boundary is now at $z=0$, and the IR hard wall is at $z=z_0$.
The Schrödinger potential $V_\mathrm{WSS}$ is given by
\begin{align}
V_\mathrm{WSS}=    \frac13 \frac{1+\frac23 Z^2}{(1+Z^2)^{2/3}},
\end{align}
where $Z=Z(z)$ is the inverse of the composition of the two coordinate transformations \eqref{eq:coordtraf1SS} and \eqref{eq:coordtraf2SS}. At small $z$, we have 
\begin{align}
\label{eq:SSsmallzpot}
    V_\mathrm{WSS}\sim\frac{2}{z^2},
\end{align}
while close to $z=z_0$,
\begin{align}
    V_\mathrm{WSS} \sim \frac13.
\end{align}
Similar to AdS models, the potential is singular near $z=0$, albeit with a different coefficient\footnote{This coefficient is determined by the behavior of $\rho$. The WSS model obtains a different coefficient since it is secretly dual to a 5-dimensional field theory.}, and regular everywhere else. One may thus solve for mode functions, bulk-to-boundary and bulk-to-bulk propagators using the WKB approximation as before. Instead of proceeding in this direction, we now prove that the HLbL scattering amplitude is finite in the WSS model.

The axial-vector contributions to the coefficient functions $\hat{\Pi}_i$ of the HLBL tensor are all of the form
\begin{align}
\label{eq:WSSPi}
     \iint_0^{z_0} dz dz' \Jpq(z,Q_1)\J(z,Q_2)\Jpq(z',Q_3) G_A(z,z',P^2),
\end{align}
where $G_A$ is the axial bulk-to-bulk propagator and $\Jpq(z,Q)={\partial_z\J(z,Q)}/{q^2}$, for some momenta $Q_i,P$. 
The integrand is a continuous function of $z,z'$ everywhere except near $z=0$ and $z'=0$ because $\J,\Jpq,G_A$ are solutions to linear differential equations\footnote{The axial bulk-to-bulk propagator has a discontinuous first derivative at $z=z'$, but is itself continuous.} whose coefficients are smooth except near $z=0$. Near $z=0$, we can determine the behavior of the solutions by using \eqref{eq:SSsmallzpot}.
The most general solution of \eqref{eq:EOMSS} at small $z$ is given by
\begin{align}
 &c_1(Q)\left( \cosh (Qz) -Qz \, \sinh (Qz)\right) \nonumber\\+&c_2(Q)\left( {\sinh (Qz)}-Qz \,\cosh (Qz) \right) \nonumber\\
 &\equiv 
 \mathcal{I}_1(z)+\mathcal{I}_2(z),
\end{align}
which is regular at $z=0$. The vector bulk-to-boundary has $c_1(Q)=1$ and $c_2(Q)$ is fixed by matching to the correct IR solution. The axial bulk-to-bulk propagator $G_A$ is proportional to 
\begin{align}
    G_A\propto \mathcal{I}_1(z)\mathcal{I}_2(z' )\theta(z'-z)+\mathcal{I}_2(z)\mathcal{I}_1(z' )\theta(z-z'),
\end{align}
and thus $\J,\Jpq,G_A$ are all continuous near $z=0$ and $z'=0$.
Thus the HLbL expressions \eqref{eq:WSSPi} are integrals of continuous functions over a compact region and therefore finite. The divergences found in the soft-wall models are not present in the WSS model.

\section{WKB approximation to the axial bulk-to-bulk propagator in the SW model}
\label{app: Btbb}
In this appendix we present WKB expressions for $ \xi_L, \, \xi_R$ which are used in the construction of the bulk-to-bulk propagator.
Due to the breakdown of the WKB approximation at very small $z<z_A=Q^{-1} (\frac{Q}{c})^a$ with $1>a>0$, we need to glue together solutions at $z_A$ as before.

We get
\begin{align}
    \xi_L(z)&=\sqrt{z} I_1(Q z) \;\;\;\;\;\;\;\;\;\;\; \; z\in [0,z_A] \\
    \xi_L(z)&= \sqrt{\frac{1}{2\pi}}\frac{e^{Q z_A}}{\sqrt{\kappa}}e^{\int_{z_A}^z \kappa }\;\;\;\; z\in [z_A,\infty)
\end{align}
and

\begin{align}
   \xi_R(z)&=\sqrt{z} K_1(Q z) \;\;\;\;\;\;\;\; \;\;\;\;  z\in [0,z_A] \\
    \xi_R(z)&= \sqrt{\frac{\pi}{2}}\frac{e^{-Q z_A}}{\sqrt{\kappa}}e^{-\int_{z_A}^z \kappa }\;\;\;\; \;z\in [z_A,\infty)
\end{align}

With this choice, $\xi_L'(z')\xi_R(z')- \xi_L(z') \xi_R'(z')=1$ for all $z'$, and the bulk-to-bulk propagator $G_A(z,z',Q^2)$ given in \eqref{eq:bulktoboundaryaxialfin} is symmetric in $z, \, z'$.
As in the discussion of the individual modes, the gluing is only smooth in the $Q\rightarrow \infty$ limit. As we are interested in integrals, rather than derivatives of such functions, these mild discontinuities at $z_A$ play no important role.

\section{Nonuniform convergence of mode sums}
\label{app:nonuniform}
In section \ref{sec:noncomm} we found two different results for the VVA correlator, specifically the coefficient function $\bar{\mathcal{W}}_2(q^2)$, depending on how the computation was performed. When evaluating the 5d Witten diagram using vector and axial vector bulk to boundary propagators a well defined and finite $\bar{\mathcal{W}}_2(q^2)$ was found, while when reconstructing it dispersively using axial TFFs a divergence was encountered in the infinite sum over resonances. 
As we will now argue, this apparent contradiction arises from an illegal exchange of limits.

We will first decompose $\mathcal{A}^\perp$ into a sum over normalizable modes. This sum over modes converges pointwise to $\mathcal{A}^\perp$, but it  cannot do so uniformly. Due to non-uniform convergence, the exchange of the infinite sum with the definite integral $\int_0^\infty dz$ over the infinite interval is not allowed, preventing 
a decomposition of $\bar{\mathcal{W}}_2$ into a sum over modes.

One may calculate the inner product of the axial bulk-to-boundary propagator with a normalizable mode to find 
\begin{align}
    \langle A^{\perp}_n, \mathcal{A}^{\perp} \rangle=-\frac{F_n^A}{Q^2+(M_n^A)^2}.
\end{align}
This equation holds for the exact modes; if one instead uses the WKB approximations, one has to expect corrections to the RHS. These corrections are, however, irrelevant at large $Q$ and $n$ and do not affect convergence issues.
Hence, 
\new{we can use these approximations to study the convergence of}
\begin{align}
\label{eq:axmodsum1}
    \mathcal{A}^{\perp}_{\text{sum}}=- \sum_n \frac{F^A_n}{Q^2+(M_n^A)^2} A^{\perp}_n(z)
\end{align}
\new{to $\mathcal{A}^{\perp}$.}
We begin our analysis by switching to Schrödinger coordinates ${A}^\perp={\xi}/{\sqrt{\rho}}$ with $\rho ={e^{-\phi}}/{z}$
and define
\begin{align}
\label{eq:axmodsum2}
      \xi^{}_{\text{sum}}=- \sum_n \frac{F^A_n}{Q^2+(M_n^A)^2} \xi^{}_n(z).
\end{align}
The individual contributions of \eqref{eq:axmodsum2} are just the individual terms of \eqref{eq:axmodsum1} multiplied by $\sqrt{\rho}$. This may seem like a trivial step, but it turns out that the two sums \eqref{eq:axmodsum1} and \eqref{eq:axmodsum2} have different convergence properties.

Let us first check whether \eqref{eq:axmodsum2} converges pointwise.
For any point\footnote{\label{footnote:zeroz}At $z=0$, the mode sum will always give $\mathcal{A}_{\text{sum}}^\perp=0$ since each mode vanishes at the origin, while the bulk to boundary propagator is 1. This is, however, just an isolated point. In any integral of $\mathcal{A}^\perp$, this will not matter.} $z \neq 0$, there will be a mode number $N$ such that for all modes with $n>N$ the point $z$ will always lie in region \textrm{II}. The left and right boundaries $z_1,z_2$ of region II behave like 
\begin{align}
    z_1&\sim E_n^{-\frac{1}{2}} \rightarrow0\\
    z_2& \sim E_n^{\frac{1}{4}}\rightarrow \infty
\end{align}
as $n\rightarrow \infty$. Hence, any given point $z$
is eventually within
region II, where
\begin{align}
  \xi_n(z)&\sim \frac{C_1^A}{\sqrt{k(z)}} \cos(\int_{z*}^z dz' k(z')+ \tilde{\varphi}_0)  \\&\sim E_n^{-\frac{1}{8}}\cos(\sqrt{E_n}z+ \varphi_0),
\end{align}
with $\varphi_0$ being an $n$-independent constant phase shift. 

In appendix \ref{app:WKBDetails}, the axial decay constant scaling is determined to be $F_n^A \sim C_1^A \sqrt{E_n} \sim E_n^{\frac{1}{8}}\sqrt{E_n}$. We may also approximate $Q^2+(M_n^A)^2\sim(M_n^A)^2=E_n$ at sufficiently high $E_n=m_n^2 \sim n^{4/3}$.

The sum \eqref{eq:axmodsum2} at high mode numbers $n$ then behaves as 
\begin{align}
    \sum_n E_n^{-1/2} \cos (\sqrt{E_n}z+\varphi_0)= \sum_n n^{-2/3} \cos (n^{2/3}z+\varphi_0)
\end{align}
When studying its convergence properties, one notices that the sequence $s_n=n^{2/3}$ becomes denser for large $n$,
\begin{align}
    s_{n+1}-s_n=\frac{2}{3} n^{-1/3}.
\end{align}
 This allows one to approximate the above sum beyond some large $n=\tilde{N}$ by the integral\footnote{We thank Harald Skarke for discussions on this point.}
 \begin{align}
     \frac{3}{2} \int_{\tilde{N}^{2/3}}^{\infty} \frac{\cos(zx+\varphi_0)}{\sqrt{x}} dx.
 \end{align}
\new{This can be integrated in closed form, where the contribution from the upper integration boundary vanishes like}
\begin{align}
\label{eq:intresult}
    \frac{3}{2} \frac{1}{z \sqrt{x}} \sin (z x+ \varphi_0)\bigg|_{x\to\infty}=0.
\end{align}
The mode sum $\xi_{\text{sum}}$ therefore converges pointwise to the full bulk-to-boundary propagator $\xi$ at any nonzero $z$, although the limit is approached quite slowly. This same argument also works for $ \mathcal{A}^{\perp}_{\text{sum}}$ in \eqref{eq:axmodsum1}.

The convergence of $\xi_{\text{sum}}$ is also uniform on any compact interval excluding zero (a small region $[0,\delta)$ around $z=0$ is excluded as explained in footnote \ref{footnote:zeroz}). We recapitulate some aspects of uniform convergence in Appendix \ref{app:uniconv}.

Uniform convergence of $\xi_{\text{sum}}$ on any compact interval excluding zero is relatively easy to show.
We need to show that 
\begin{align}
h_N(z)&=\xi(z)-\sum_{n=0}^{N-1} \frac{-F_n^A}{Q^2+(M_n^A)^2}\xi_n(z)\nonumber\\&=\sum_{n=N}^\infty \frac{-F_n^A}{Q^2+(M_n^A)^2}\xi_n(z)
\end{align}
is bounded by a sequence $b_N$ on the compact interval, where $b_N$ goes to zero as $N\rightarrow \infty$. For any compact interval, there exists a mode number $\hat{N}$, such that the interval is completely contained in region II of the $\hat{N}$'th WKB solution. It will also be contained in region II of all WKB solutions with $N > \hat{N}$. Assuming $N > \hat{N}$ and big enough so that the approximation of the sum by an integral works as before, one can estimate
\begin{align}
    h_N(z)\sim \frac{2}{3}\frac{1}{z\sqrt{N}}\sin(z N+\varphi_0)
\end{align}
whose absolute value is bounded by $b_N=\frac{b}{\sqrt{N}}$, where $b$ depends on the size of the interval.

Unfortunately, it seems technically difficult to prove uniform convergence on the whole infinite interval $[\delta,\infty)$ even though it seems likely. At large $z$, not all of the $\xi_n$ are described by the oscillating WKB solution of region II, but rather the expressions for regions III and IV. Out of the infinite sum  $\sum_{n=N}^\infty$ only a finite number of terms are described by regions III and IV at fixed $z$, but that number grows as $z$ grows.

\begin{figure}[h]
    \centering
    \includegraphics[width=1\linewidth]{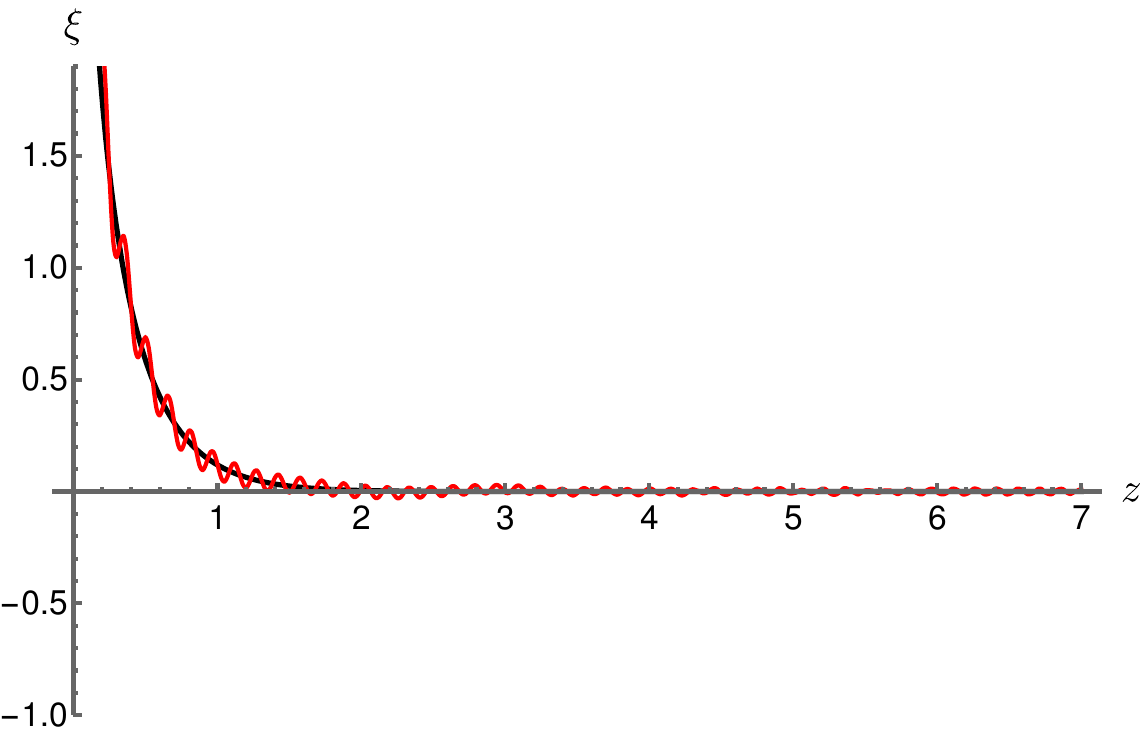}
     \includegraphics[width=1\linewidth]{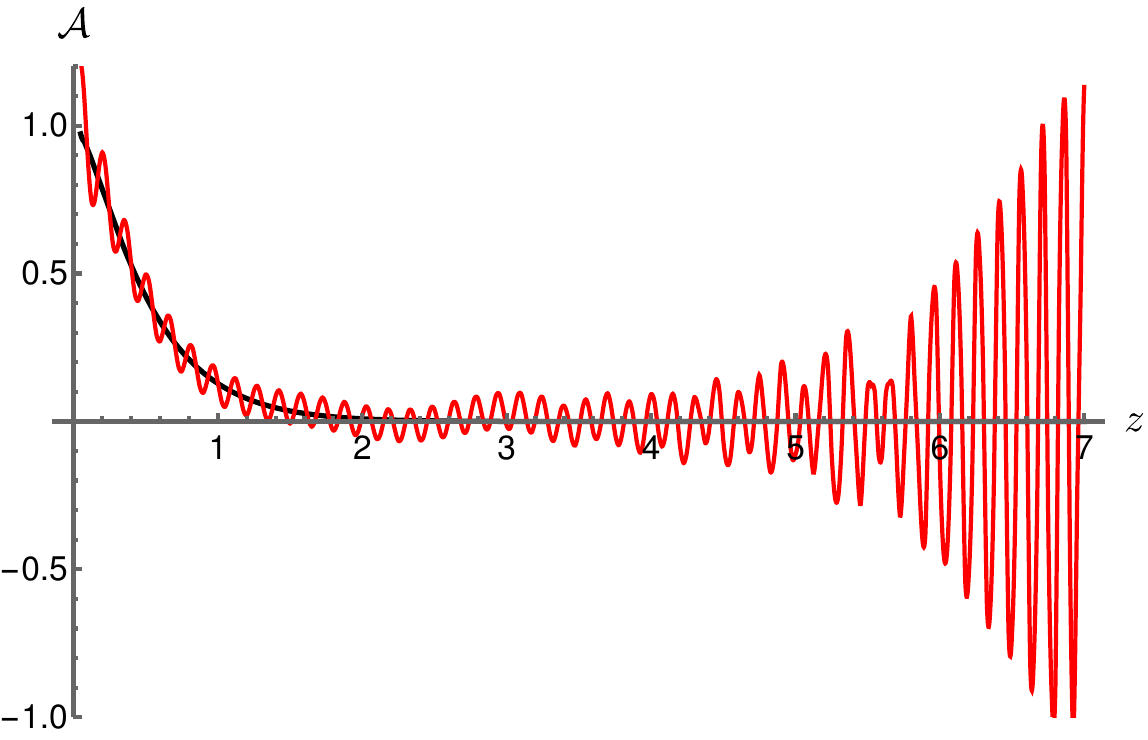}
    \caption{The mode sum (red) compared to the bulk-to boundary propagator (black) for the first 1400 modes at $Q=3\; \rm GeV$ {as a function of $z$ (in units of GeV$^{-1}$). $\xi$ is plotted in units of $\rm GeV^{-1/2}$, while $\mathcal{A}$ is dimensionless.}
    On the upper plot, the convergence to $\xi$ (which goes like $\frac{1}{\sqrt{z}}$ at small $z$) is uniform. Upon multiplying by $\frac{1}{\sqrt{\rho}}$, one obtains the lower plot, where the convergence is not uniform anymore. The differences between the finite sum and the full solution, which shrink for growing $z$ in the Schrödinger coordinates $\xi$, get blown up at large $z$ when multiplying by $\frac{1}{\sqrt{\rho}}$.}
    \label{fig:axbtbF}
\end{figure}

Despite the numerical evidence for uniform convergence shown in the upper plot of fig. \ref{fig:axbtbF}, we were not able to prove it due to the terms that are described with the expressions of region III and IV.

The sequence $ \mathcal{A}^{\perp}_{\text{sum}}$ does, however, \emph{not} converge uniformly on the infinite interval $(\delta, \infty)$ (it does on all compact intervals excluding 0).

It is this loss of uniform convergence that is responsible for the divergence of $\sum_n \bar{\mathcal{W}}_2^{(n)}$ and which in fact provides an indirect proof of non-uniform convergence. After all, if $\mathcal{A}^\perp_{\text{sum}}$ would converge uniformly, one could insert the mode decomposition into the well-defined and finite \eqref{eq:omegaT} and interchange the infinite integral $\int_0^\infty dz$ with the infinite sum $\sum_n$. 

Uniform convergence would allow us to interchange these two limits without a penalty, but clearly the results are very different in our case depending on which of the two operations we perform first. Using the exponential scaling of the TFFs we showed before that if the mode sum is performed last, the result is ill-defined, whereas if one performs the mode sum first, the result is finite. So uniform convergence cannot hold.

This proof, which used the exponential scaling of the TFFs, is rather indirect and a more direct proof using the expressions from the WKB approximation would be desirable but is lacking at the moment. The bottom plot of fig. \ref{fig:axbtbF} exhibits numerical evidence for non-uniform convergence.

It may seem puzzling that $\xi_{\text{sum}}$ converges uniformly on $(\delta, \infty)$, but $\mathcal{A}^\perp_{\text{sum}}$ does not since the individual summands just differ by multiplication by $\frac{1}{\sqrt{\rho}}= e^{+\frac{c^2 z^2}{2}} \sqrt{z}$. We explain how this can come about with a simple example in appendix \ref{app:uniconv}.

Previously, we also argued that the sum rule for pions
\begin{align}
\label{eq:sumrulen}
    \sum_n F_n(0,0)f_n =\frac{N_c}{2 \pi^2}\text{tr}(t^3 \Q^2  ).
\end{align}
does not converge. Let us try to approach this problem from the 5-dimensional theory first and see where things go wrong.
With $\Phi=(\varphi,\eta)$, the inner product between a putative non-normalizable mode $\Phi_S$ (which obeys the same boundary conditions at $z=\infty$ as the normalizable mode) and a normalizable mode reduces to a boundary term and if we choose $\varphi \rightarrow 0,\, \eta \rightarrow 1$ at $z=0$ for the non-normalizable mode, we get\footnote{Note that these boundary conditions are different from the ones used in the VVA correlator.}
\begin{align}
\label{eq:phimodesum}
    \langle \Phi_n, \Phi_S \rangle =-\frac{m_n^2}{q^2-m_n^2}\frac{\varphi_n'(z)}{z g_5^2}\bigg|_{z=0}= \frac{m_n^2 f_n}{q^2-m_n^2}.
\end{align}
We will mostly be working at Euclidean $q^2=-Q^2<0$.
Using the WKB approximation, one can try to see if such a non-normalizable solution exists. For this investigation, we need to work away from the chiral limit, as in the discussion of the pion decay constants before.
We first need to translate the boundary conditions at $z=0$ for $(\varphi, \eta)$ to $y$ in the Schrödinger form.
For Euclidean $q^2=-Q^2$ and nonzero quark mass, one can find an analytic solution for small $z$
\begin{align}
    y=\bar{C}_1 \sqrt{z}K_0(Qz)
\end{align}
The other linearly independent solution $\sqrt{z}I_0(Qz)$ is exponentially rising for large $Qz$ and cannot be matched to the WKB solution.
The boundary condition at small $z$ reads $\partial_z\left( {y}/{z} \right)={m_q g_5}/{z}$, which implies $\bar{C}_1=-m_q g_5$.

The WKB solution that falls off at infinity is
\begin{align}
    y=\frac{\bar{C}_2}{\sqrt{\kappa}}\exp\left[-\int_{z_A}^{z} dz'\,
    \kappa(z')\right].
\end{align}
Matching the two solutions, one has $\bar{C}_2= \bar{C}_1 e^{-Q z_A}\sqrt{Q} \sqrt{\frac{\pi}{2}}$. The point $z_A$ scales in  the same way as in the axial-vector case.
The non-normalizable mode can therefore be explicitly constructed with the WKB method.
As in the axial-vector case one may attempt to decompose $y(z)$ in terms of modes 
\begin{align}
\label{eq:ypseudosumrule}
    y_{\text{sum}}(z)= \sum_n^{\infty}\frac{m_n^2}{q^2-m_n^2}f_n y_n(z)
\end{align}
One can ask again if this expression converges, and if the answer is yes, what kind of convergence one has.
As the analysis is very similar to the axial case we will not go into full detail for the pseudoscalars.
As before, if we fix an arbitrary point $z$, there exists an $N$ such that for all $n>N$, $z$ will lie in region \textrm{II}.
Then at any point 
\begin{align}
    m_q \sum_n n^{-4/3} \cos(n^{2/3}z+ \tilde{\varphi}_0)
\end{align}
has to converge in order for \eqref{eq:ypseudosumrule} to converge pointwise which is indeed the case\footnote{Due to the sequence $s_n=n^{2/3}$ getting denser for higher $n$, the sum may be replaced by the integral $\int dx \cos(zx+ \tilde{\varphi}_0) x^{-5/3}$ similar to the axial case before.}. As before, uniform convergence on the whole domain except a small region around $z=0$ seems to hold, although we have no proof.

When expressing this sum rule in terms of $\partial_z\varphi$ we have to divide both sides of \eqref{eq:ypseudosumrule} by $\sqrt{\omega} \frac{e^{-\phi}}{z} \sim \frac{z^{-\frac{5}{2}}}{g_5 \sigma} e^{- \frac{c^2 z^2}{2}}$, which makes the convergence non-uniform.

Therefore, when integrating the equation
\begin{align}
\label{eq:phiprisumrule}
    \partial_z\varphi(z)= \sum_n^\infty\frac{m_n^2}{q^2-m_n^2}f_n \partial_z \varphi_n(z)
\end{align}
with $\int_0^\infty dz$, it is forbidden to interchange the sum and the integral. The LHS is completely harmless, giving always\footnote{Away from the chiral limit, as is necessary in the sum rule, one has $\varphi(\infty)=0$ for all $Q^2$.} $- \varphi(0)=-1$, while if we were to exchange the integral with the sum, the RHS would read $\sum_{n=1}^\infty \frac{m_n^2}{q^2-m_n^2}f_n \frac{12 \pi^2}{N_c}F_n(0,0)$ which diverges due to the exponential scaling of the TFF.

 Hence $\varphi(\infty) =\int \partial_z \varphi$ is perfectly well defined, but it is not computed by $\lim_{N\rightarrow \infty}\int \sum_n^N(...)$ as this limit does not exist. The pseudoscalar sum rule \eqref{eq:sumrulen} derived for hard-wall models simply does not exist in soft-wall models.

\section{Uniform convergence}
\label{app:uniconv}
Uniform convergence of a sequence of functions $f_i$ to a limit $f$ on an interval $I$ means that given any $\varepsilon>0$, one can find an $N\in \mathbb{N}$ such that $\forall n>N$ and $z \in I$,
\begin{align}
    \big|f(z)-\sum_{i=1}^n f_i(z)\big|< \varepsilon.
\end{align}
Basically, it means for $h_n(z)=f(z)-\sum_{i=1}^n f_i(z)$ that its absolute value is bounded by a sequence of constants on \emph{all of $I$} which converges to zero.

In Appendix \ref{app:nonuniform},
we claimed that $\xi_{\text{sum}}$ converges uniformly on $[\delta, \infty]$, while $\mathcal{A}^\perp_{\text{sum}}$ does not.
Here we would like to illustrate that this is possible by a simple example. Pick a sequence of functions $f_n(z)=\frac{1}{n} \theta\left(z-(z_n-\Delta)\right) \theta\left(-z+(z_n +\Delta)\right)$with $z_n$ a monotonic sequence that goes to infinity. This sequence describes rectangular bumps of fixed length that move to the right and whose height decreases as $n$ increases. Clearly $f_n$ converges to $0$ both pointwise and uniformly since $h_n$ is a bounded function whose bound goes to zero as $n\rightarrow \infty$.

When we multiply $f_n$ by $e^{c^2 z^2}$, the new sequence will, in general, \emph{not} converge to $e^{c^2 z^2} \times 0=0 $ uniformly on all of $\mathbb{R}$, even though it does converge pointwise. Upon picking $z_n = n$, the sequence $e^{c^2 z^2}f_n$ is now given by a function which is again only nonzero in $[n- \Delta,n+ \Delta]$, but its average height is $\frac{1}{n} e^{c^2 n^2}$ which is unbounded. Hence, the convergence is not uniform on $(0, \infty)$ in this simple example. Note that on any \emph{finite} interval $[a,b]$ the convergence is still uniform, only on the infinite interval $(0, \infty)$ it is not.

Also in this case we have a non-commutativity of limits.
If we were to integrate the limit function over the infinite interval $[0,\infty)$, then we would get zero, as the pointwise limit is the zero function.
Integrating $e^{c^2z^2}f_n(z)$, one would get something of the order of $2 \Delta \frac{1}{n} e^{c^2 n^2}$ whose limit does not exist. The integration $\int _0^\infty dz$ and the limit $\lim_{n\rightarrow \infty}$ cannot be exchanged in this simple example.

\bibliographystyle{elsarticle-num}
\bibliography{SM,hlbl+}

\end{document}